\documentclass[12pt]{article}
\usepackage{epsf}
\title{ {\bf
Leading logarithmic QCD corrections to the $B_{s}\rightarrow 
\gamma\gamma$ decays in the two Higgs doublet model}}
\author{\vspace{1cm}\\
         {\bf T. M. Aliev} , \\
        International Centre for Theoretical Physics \\
        Trieste, Italy \\
        and \\          
        Physics Department, Girne American University \\
        Girne , Cyprus \\
        \vspace{5mm}\\
        {\bf G. Hiller}\thanks{E-mail address: ghiller@x4u2.desy.de} , \\
         Deutsches Elektronen-Synchrotron DESY, Hamburg \\
        \vspace{5mm}\\
        {\bf E. O. Iltan}
        \thanks{E-mail address:
        eiltan@heraklit.physics.metu.edu.tr}
 \\
        Physics Department, Middle East Technical University \\
        Ankara, Turkey\\}

\date{}

\begin{document}
\setlength{\baselineskip}{24pt}

\maketitle
\setlength{\baselineskip}{7mm}
\begin{abstract}
We calculate the leading logarithmic QCD corrections to the decay
$B_{s}\rightarrow\gamma\gamma$ in the two Higgs doublet model (2HDM)
including $O_{7}$ type long distance effects
and estimate the restrictions of the 2HDM
parameters, $tan\beta$ and $m_{H}$, using the experimental data
of $B \rightarrow X_s \gamma$ decay provided by the CLEO Collaboration.
A lower bound for the charged Higgs mass $m_H$ as a function of the 
renormalization scale $\mu$ is given for 2HDM model II.
We further present the dependencies of the branching ratio
$Br(B_{s}\rightarrow\gamma\gamma)$ and 
the ratio $|A^{+}|^2/|A^{-}|^2$
on $m_{H}$ and $tan\beta$ including leading logarithmic QCD
corrections. The dependence on the renormalization scale is found to be
strong for both ratios. 
An additional uncertainty arises from the variation of the parameters of the 
bound state model, $(m_b, \bar{\Lambda}_s)$.
We see, that to look for charged Higgs effects the measurement of the 
branching ratio $Br(B_{s}\rightarrow\gamma\gamma)$ is promising.
\end{abstract} 
\thispagestyle{empty}
\newpage
\setcounter{page}{1}

\section{Introduction}
The experimental discovery of the inclusive and exclusive $B\rightarrow X_s
\gamma$ \cite{cleo} and $B\rightarrow K^* \gamma $ \cite{Rammar} 
decays stimulated the study
of rare B meson decays as a new force. These decays take place via
flavor-changing neutral current (FCNC) $b \rightarrow s$ transitions, which
are absent in the Standard Model (SM) at tree level and appear only 
through loops. Therefore, the study of rare B-decays can provide
a sensitive test of the structure of the SM at loop
level and 
may shed light on the Kobayashi-Cabibbo-Maskawa
(CKM) matrix elements and the
leptonic decay constants of the B-mesons. At the same time, such
decays are in a very promising class to search for new
physics beyond the SM, like two Higgs doublet model (2HDM),
minimal supersymmetric model (MSSM), etc. \cite{Hewett}.
Currently, the main interest is focused on these 
decays for which the SM predicts large branching ratio and which can be
measured in near future in the constructed B-factories. The $
B_s\rightarrow \gamma\gamma$ decay belongs to this category. In the 
SM, the branching ratio of $B_s\rightarrow \gamma\gamma$
decay is of order  $10^{-7}$ without QCD corrections.
Including leading log (LLog) QCD corrections, the branching ratio $(Br)$
$b\rightarrow s\gamma\gamma$ is of the same order of magnitude 
$\sim 10^{-6}$ like $b\rightarrow s l^{+} l^{-}$ \cite{morozumi}.

The investigation of $B_s\rightarrow \gamma\gamma$
decay is interesting for the following reasons:
\begin{itemize}
\item It is well-known, that the QCD corrections to the 
$b\rightarrow s\gamma$ 
decay are considerably large (see \cite{Grinstein} - \cite{burasmisiak} 
and references therein).
Therefore, one can naturally expect that the situation is the same 
for the $b\rightarrow s \gamma\gamma$ decay. 
Recently the QCD corrections in the 
LLog approximation to this decay have been calculated and found
to be large as expected \cite {Gud} - \cite {Soni}.
Note, that in the literature this decay without QCD corrections was 
analysed in the SM \cite{yao} - \cite{simma} and in the 2HDM \cite{aliev}. 

\item  In $B_s\rightarrow \gamma\gamma$ decay, the final photons can be in
a CP-odd or a CP-even state. Therefore this decay allows us to study  CP
violating effects.

\item  From the experimental point of view, $B_s\rightarrow \gamma\gamma$ 
decay  can be easily identified by putting
a cut for the energy of the final photons, e.g., the energy of each
photon is larger than $100\,\, MeV$. In this case, two hard photons
will be easily detected in the experiments \cite{Riccardi}.

\item  Finally, this decay is also sensitive to the physics beyond the SM.
\end{itemize} 
In an earlier analysis \cite{aliev}, the 
$Br (B_s\rightarrow \gamma\gamma)$ in the 2HDM without QCD
corrections was found to be enhanced with respect to the SM one for some 
values of the parameter space.
In the present work, we study $B_s\rightarrow \gamma\gamma$ decay in
the 2HDM with perturbative QCD
corrections in LLog approximation. 
In contrast to \cite{aliev}, who used the constituent quark model, we 
impose a model based on 
heavy quark effective theory for the bound state of the meson $B_s$.
Further we perform an additional analysis with the inclusion of 
long-distance effects through the transition 
$B_s \to \phi \gamma \to \gamma \gamma$, which we call $O_7$-type throughout 
this paper, see \cite{Gud} for details. 
We find, that the theoretical 
analysis is shadowed by large uncertainties due to the renormalization 
scale $\mu$ and the parameters of the bound state.
The decay $B_s \to \gamma \gamma$ is dominated by the Wilson coefficient
$C_7^{eff}$ (see section 2), which is restricted in our analysis by the
$B \to X_s \gamma$ branching ratio provided by CLEO data \cite{cleo},
$Br(B \to X_s \gamma) \propto |C_7^{eff}|^2$, see section 3.
Without any improvement from the theoretical side, we see that
the only chance to detect a deviation from the SM in 
$B_{s}\rightarrow\gamma\gamma$ decay lies in a possible 
enhanced branching ratio,
which can be at most $1.4 \cdot 10^{-6}$ in the SM \cite{Gud} and 
$2.1 \cdot 10^{-6}$ in model II of the 2HDM 
(for $m_H=480$ GeV and large $tan\beta$)
resulting from our analysis, at $\mu=2.5$ GeV 
including the $O_7$-type long distance effects. 

The paper is organized as follows:
In Section 2, we give the LLog QCD corrected Hamiltonian responsible for
the $b\rightarrow s\gamma\gamma$ decay. We further 
calculate the   CP-odd $A^-$ and CP-even $A^+$ amplitudes
in an approach based on heavy quark effective theory,
taking the LLog QCD corrections into account.
In Section 3, we study the constraint analysis for the 2HDM parameters 
$m_H$ and $tan
\beta$, using the measured data on the branching ratio of the
$B \rightarrow X_{s}\gamma$ decay \cite{cleo}. 
Section 4 is devoted to an analysis of the
dependence of the ratio $|A^{+}|^2/|A^{-}|^2$ and the $Br$
on the parameters $\mu$ (scale parameter), $tan \beta$ and $m_H$ and
our conclusions.
 
\section{\bf Leading logarithmic improved short-distance contributions in 
the 2HDM for the decay $B_{s}\rightarrow \gamma \gamma$ }
Before discussing the LLog  QCD corrections to the 
$B_{s}\rightarrow \gamma \gamma$ decay, we would like to remind the main
features of the models which we use in further discussions. In the current
literature, mainly two types of 2HDM are discussed. In the so-called model I,
the up and down 
quarks get a mass via the vacuum expectation value (v.e.v) of only one
Higgs field. In model II, 
the up and down quarks get mass via v.e.v of the Higgs fields
$H_{1}$ and $H_{2}$, respectively, where $H_{1}$ ($H_{2}$) corresponds to  
first (second) Higgs doublet of the 2HDM. Note, that in this sense 
the Higgs sector of model II coincides with the MSSM extension of the SM.
In the 2HDM, there exist five physical Higgs fields, namely,
two charged $H^{\pm}$ and three neutral Higgs bosons. 
The interaction Lagrangian of the quarks with the charged fields, which we
need for the calculation of the 
$b\rightarrow s \gamma\gamma$ decay amplitude, is 
\cite{Abott}
\begin{eqnarray}
{\cal{L}}=\sqrt\frac{4 G_{F}}{\sqrt{2}} [m_{u_{i}} \xi \bar{u}_{i} L d_{j}-
m_{d_{i}} \xi ' \bar{u}_{i} L d_{j}] V_{ij} H^{+} + h.c.\,\, ,
\label{lagrangian}
\end{eqnarray}
where $L$ and $R$ denote chiral projections $L(R)=1/2(1\mp \gamma_5)$ and 
$\xi$ and 
$\xi '$ are the ratios of the two vacuum expectation values, $v_{1}$ and 
$v_{2}$ of the Higgs fields $H_{1}$ and $H_{2}$, respectively. $V_{i j}$
are the elements of the CKM matrix. 
In model II, 
\begin{eqnarray}
\xi '=-1/\xi=-tan\beta=-v_{1}/v_{2}\,\, ,
\label{modII}
\end{eqnarray}
and in model I
\begin{eqnarray}
\xi '=\xi=cot\beta=v_{2}/v_{1}.
\label{modI}
\end{eqnarray}

After this preliminary remark, let us discuss the LLog  
QCD corrections to the $b\rightarrow s \gamma\gamma $ decay amplitude 
in the 2HDM. The framework to incorporate short-distance QCD corrections in
a systematic way is that of an effective low energy theory with five quarks,
namely u,d,s,c,b quarks. The effective Hamiltonian is obtained by integrating
out the heavier degrees of freedom, i.e. the top quark, 
$W^{\pm}$ and $H^{\pm}$ bosons. In the effective
theory, only the lowest (mass) dimension operators, which are constructed
by quark and gauge fields, are taken into account,
since higher dimensional operators are suppressed by 
factors $O(m_{b}^2/m_{t}^2)$
and $O(m_{b}^2/m_{W}^2)$. 

The LLog QCD corrections are done through matching the full
theory with the effective theory at the high scale $\mu=m_W$ and then
evaluating the Wilson coefficients from $m_W$ down to the lower scale 
$\mu\sim O(m_b)$.
In this way the LLog QCD corrections for the 
$b\rightarrow s\gamma \gamma$ decay in the SM are calculated in
\cite{Gud} - \cite{Soni}.

The effective Hamiltonian relevant for our process is
\begin{eqnarray}
{\cal{H}}_{eff}=-4 \frac{G_{F}}{\sqrt{2}} V_{tb} V^{*}_{ts} 
\sum_{i=1}^{8} C_{i}(\mu) O_{i}(\mu) \, \, ,
\label{hamilton}
\end{eqnarray}
where the $O_{i}$ are operators given in eq.~(\ref{o7}) 
and the $C_{i}$ are Wilson coefficients
renormalized at the scale $\mu$. The coefficients can be calculated 
perturbatively and the hadronic matrix elements $ <V|O_{i}|B>$ can be calculated
using some non-perturbative methods.

The operator basis of ${\cal{H}}_{eff}$ is given as
\begin{eqnarray}
 O_1 &=& (\bar{s}_{L \alpha} \gamma_\mu b_{L \alpha})
               (\bar{c}_{L \beta} \gamma^\mu c_{L \beta}), \nonumber   \\
 O_2 &=& (\bar{s}_{L \alpha} \gamma_\mu b_{L \beta})
               (\bar{c}_{L \beta} \gamma^\mu c_{L \alpha}),  \nonumber   \\
 O_3 &=& (\bar{s}_{L \alpha} \gamma_\mu b_{L \alpha})
               \sum_{q=u,d,s,c,b}
               (\bar{q}_{L \beta} \gamma^\mu q_{L \beta}),  \nonumber   \\
 O_4 &=& (\bar{s}_{L \alpha} \gamma_\mu b_{L \beta})
                \sum_{q=u,d,s,c,b}
               (\bar{q}_{L \beta} \gamma^\mu q_{L \alpha}),   \nonumber  \\
 O_5 &=& (\bar{s}_{L \alpha} \gamma_\mu b_{L \alpha})
               \sum_{q=u,d,s,c,b}
               (\bar{q}_{R \beta} \gamma^\mu q_{R \beta}),   \nonumber  \\
 O_6 &=& (\bar{s}_{L \alpha} \gamma_\mu b_{L \beta})
                \sum_{q=u,d,s,c,b}
               (\bar{q}_{R \beta} \gamma^\mu q_{R \alpha}),  \nonumber   \\  
 O_7 &=& \frac{e}{16 \pi^2}
          \bar{s}_{\alpha} \sigma_{\mu \nu} (m_b R + m_s L) b_{\alpha}
                {\cal{F}}^{\mu \nu},                             \nonumber       \\
 O_8 &=& \frac{g}{16 \pi^2}
    \bar{s}_{\alpha} T_{\alpha \beta}^a \sigma_{\mu \nu} (m_b R + m_s L)  
          b_{\beta} {\cal{G}}^{a \mu \nu},  
\label{o7}
\end{eqnarray}
where  
$\alpha$ and $\beta$ are $SU(3)$ colour indices and
${\cal{F}}^{\mu \nu}$ and ${\cal{G}}^{\mu \nu}$
are the field strength tensors of the electromagnetic and strong
interactions, respectively.

For the reason given below, the LLog QCD corrections  can be 
calculated in analog to the SM.
In the 2HDM, the charged Higgs fields are present and 
give new contributions due to the their exchange diagrams.
Since the interaction vertices 
of the charged Higgs bosons and quarks are proportional to the
ratio $m_{q}/m_{W}$, where $m_{q}$ is the mass of the quark and $m_{W}$ is
the mass of the W boson, the main contribution comes from the interaction with
the t-quark.
We neglect the contributions coming from $u$ and $c$ quarks, 
since their masses are negligibly small compared to $m_{W}$. 
In this case the calculations show, that the new contributions 
modify only the 
Wilson coefficients $C_{7}$ and $C_{8}$ of the 
operators $O_{7}$ and $O_{8}$ at $m_{W}$ scale and do not bring any new
operators \cite{aliev}. 
Therefore the operator basis used in the 2HDM is the same as the basis
used in the SM for the $b\rightarrow s \gamma\gamma$ decay.

Denoting the Wilson coefficients for the SM with $C_{i}^{SM}(m_{W})$ and the
additional charged Higgs contribution with $C_{i}^{H}(m_{W})$, 
we have the initial values \cite{effham}:
\begin{eqnarray}
C^{SM}_{1,3,\dots 6}(m_W)&=&0 \nonumber \, \, , \\
C^{SM}_2(m_W)&=&1 \nonumber \, \, , \\
C_7^{SM}(m_W)&=&\frac{3 x^3-2 x^2}{4(x-1)^4} \ln x+
\frac{-8x^3-5 x^2+7 x}{24 (x-1)^3} \nonumber \, \, , \\
C_8^{SM}(m_W)&=&-\frac{3 x^2}{4(x-1)^4} \ln x+
\frac{-x^3+5 x^2+2 x}{8 (x-1)^3}\,\, ,
\label{Coeffsm}
\end{eqnarray}
and 
\cite{aliev}, \cite{Grinsteinsav}
\begin{eqnarray}
C^{H}_{1,\dots 6}(m_W)&=&0 \nonumber \, \, , \\
C_7^{H}(m_W)&=&\xi \xi' [\frac{-3 y^2+2 y}{6(y-1)^3} \ln y+
\frac{3 y-5 y^2}{12 (y-1)^2}] \nonumber \, \, , \\ 
&+& \xi^{2} [\frac{3 y^3-2 y^2}{12 (y-1)^4} \ln y+
\frac{-8 y^{3}-5 y^2+7 y}{72 (y-1)^3}] \nonumber \, \, , \\
C_8^{H}(m_W)&=&\xi \xi' [\frac{y}{2(y-1)^3} \ln y+
\frac{y^2-3 y}{4 (y-1)^2}] \nonumber \, \, , \\ 
&+& \xi^{2} [\frac{y^2}{4 (y-1)^4} \ln y+
\frac{-y^{3}+5 y^2+2 y}{24 (y-1)^3}]  \, \, , 
\label{CoeffH}
\end{eqnarray}
where $x=m_t^2/m_W^2$ and $y=m_t^2/m_H^2$.
Here the parameters $\xi$ and $\xi'$ are given in eqs.~(\ref{modII}) and 
~(\ref{modI}).
From eqs.~(\ref{Coeffsm}) and (\ref{CoeffH}) the initial values of
the coefficients for the 2HDM are defined as: 
\begin{eqnarray}
C^{2HDM}_{1,3,\dots 6}(m_W)&=&0 \nonumber \, \, , \\
C_2^{2HDM}(m_W)&=&1 \nonumber \, \, , \\
C_7^{2HDM}(m_W)&=&C_7^{SM}(m_W)+C_7^{H}(m_W) \nonumber \, \, , \\
C_8^{2HDM}(m_W)&=&C_8^{SM}(m_W)+C_8^{H}(m_W). 
\label{Coef2HDM}
\end{eqnarray}

Using the initial values of the Wilson 
coefficients $C_i^{2HDM}$, we can calculate their contributions at any 
lower scale as in the SM case. Here we would like to 
make the following remark: Since in our case 
there exists a charged Higgs boson with a mass 
larger than $m_{W}$, the correct procedure to
calculate the Wilson coefficients at a lower scale $\mu$ has two 
stages: First, we calculate the value at $m_W$ 
starting from $m_H$ and second, 
we evaluate the result from $m_W$ down to a lower scale $\mu$.  
We assume that the evaluation from $m_{H}$ to $m_{W}$ gives 
a negligible contribution to the Wilson coefficients 
and therefore we consider only their evaluation between $m_W$ 
and a lower scale $\mu$.

Using the effective Hamiltonian in eq.~(\ref{hamilton}),
the amplitude for the decay $B_{s}\rightarrow \gamma \gamma$ can be 
written as
\cite{Gud} - \cite{aliev}
\begin{equation}
{\cal A}(B_{s}\rightarrow \gamma \gamma)=
A^{+} {\cal F}_{\mu\nu} {\cal F}^{\mu\nu} +
i A^{-} {\cal F}_{\mu\nu} \tilde{{\cal F}}^{\mu\nu}
\, \, ,
\label{amp}
\end{equation}
where 
$\tilde{{\cal F}}_{\mu\nu}=\frac{1}{2}\epsilon_{\mu\nu\alpha\beta} 
{\cal F}^{\alpha\beta}$.
Here $A^{+}$ ($A^{-}$) is the CP-even (CP-odd) part in a HQET inspired 
approach \cite{Gud} :
\begin{eqnarray}
A^{+}&=&\frac{\alpha_{em} G_F}{\sqrt{2} \pi} \frac{f_{B_s}}{m_{B_{s}}^{2}} 
\lambda_t 
\left( \frac{1}{3}
\frac{m^4_{B_s} (m_b^{eff}-m_s^{eff})}{\bar{\Lambda}_s 
(m_{B_s}-\bar{\Lambda}_s) (m_b^{eff}+m_s^{eff})} 
C_7^{eff}(\mu)
\nonumber
\right.\\
&-&
\left.
\frac{4}{9} \frac{m_{B_{s}^2}}{m_b^{eff}+m_s^{eff}}
(-m_b J(m_b)+ m_s J(m_s) ) D(\mu) 
\right)  ,\nonumber\\
A^{-}&=&- \frac{\alpha_{em} G_F}{\sqrt{2} \pi} f_{B_s} \lambda_t 
\left( \frac{1}{3}
\frac{1}{m_{B_s} \bar{\Lambda}_s (m_{B_s}-\bar{\Lambda}_s)} g_{-}
C_7^{eff}(\mu) \nonumber
\right.\\
&-&
\left.
\sum_q Q_q^2 I(m_q) C_q(\mu) +
\frac{1}{9 (m_b^{eff}+m_s^{eff})} 
(m_b \triangle(m_b)+m_s \triangle(m_s)) D(\mu)
\right) \, \, ,
\label{Amplitudes}
\end{eqnarray}
where $Q_q=\frac{2}{3}$ for $q=u,c$ and $Q_q=-\frac{1}{3}$ for $q=d,s,b$.
In the calculations, we have used the unitarity of the CKM-matrix 
$\sum_{i=u,c,t} V_{is}^{*} V_{ib}=0 $ 
and have neglected the contribution due to 
$V_{us}^{*} V_{ub} \ll V_{ts}^{*} V_{tb}\equiv \lambda_t$. 
The function $g_{-}$ is defined as:
\begin{eqnarray}
g_{-}=m_{B_s}(m_b^{eff}+m_s^{eff})^2+
\bar{\Lambda}_s (m^2_{B_s}-(m_b^{eff}+m_s^{eff})^2) 
\,\, .
\end{eqnarray}
The parameter $\bar{\Lambda}_{s}$ enters eq.~(\ref{Amplitudes}) through the
bound state kinematics \cite{Gud}.
$m_b^{eff}$ and $m_s^{eff}$ are the effective masses
of the quarks in the $B_{s}$ meson bound state \cite{Gud},
\begin{eqnarray}
(m_{b}^{eff})^{2}&=& p^2=m_{b}^{2}-3 \lambda_{2}\,\, , \nonumber \\
(m_{s}^{eff})^{2}&=& p'^2=(m_{s}^{eff})^{2}-m_{B_{s}}^{2}+
2 m_{B_{s}}\bar{\Lambda}_{s}\,\, , 
\label{HQET}
\end{eqnarray}
where $\lambda_{2}$ comes from the matrix element of the heavy
quark expansion \cite{MW}.
The LLog QCD-corrected Wilson coefficients 
$C_{1 \dots 6}(\mu)$ \cite{Gud} - \cite{Soni}
enter the amplitudes in the combinations
\begin{eqnarray}
C_u(\mu)&=&C_d(\mu)=(C_3(\mu)-C_5(\mu)) N_c +C_4(\mu)-C_6(\mu) \, \, , 
\nonumber \\
C_c(\mu)&=&
(C_1(\mu)+C_3(\mu)-C_5(\mu)) N_c +C_2(\mu)+C_4(\mu)-C_6(\mu) \, \, ,
\nonumber \\
C_s(\mu)&=&C_b(\mu)=
(C_3(\mu)+C_4(\mu))(N_c+1)-N_c C_5(\mu)-C_6(\mu) \, \, , \nonumber \\
D(\mu)&=&C_5(\mu)+C_6(\mu) N_c \, \, ,
\end{eqnarray} 
where $N_c$ is the number of colours ($N_c=3$ for QCD).
While $C_{1 \dots 6}(\mu)$ are the coefficients of the operators 
$O_{1 \dots 6}$, $C_7^{eff}(\mu)$ is the "effective" coefficient of $O_7$ and 
contains renormalization scheme dependent contributions from the 
four-quark operators
$O_{1\ldots 6}$ in ${\cal H}_{eff}$ to the effective vertex in
$b \rightarrow s \gamma$.
In the NDR scheme, which we use here, 
$C_7^{eff}(\mu)=C_7(\mu)-\frac{1}{3} C_5(\mu)-C_6(\mu)$, see \cite{effham} 
for details.
The functions $I(m_q), \, J(m_q)$ and $\triangle(m_q)$ come from the 
irreducible diagrams with an internal 
$q$ type quark propagating and are defined as
\begin{eqnarray}
I(m_q)&=&1+\frac{m_q^2}{m_{B_s}^2} \triangle (m_q) \, \, , \nonumber \\
J(m_q)&=&1-\frac{m_{B_s}^2-4 m_q^2}{4 m_{B_s}^2} \triangle(m_q)  \, \, ,
\nonumber \\
\triangle(m_q)&=&\left(
\ln(\frac{m_{B_s}+\sqrt{m_{B_s}^2-4 m_q^2}}
         {m_{B_s}-\sqrt{m_{B_s}^2-4 m_q^2}})-i \pi \right)^2 
\, \,{\mbox{for}}\, \, \frac{m^2_{B_s}}{4 m_q^2} \geq 1 , \nonumber \\
\triangle(m_q)&=&-\left(
2 \arctan(\frac{\sqrt{4 m_q^2-m_{B_s}^2}}
         {m_{B_s}})-\pi \right)^2 
\, \,{\mbox{for}}\, \, \frac{m^2_{B_s}}{4 m_q^2} <\ 1 .
\end{eqnarray}
In our numerical analysis we used the input values given in 
Table~(\ref{input}).
\begin{table}[h]
        \begin{center}
        \begin{tabular}{|l|l|}
        \hline
        \multicolumn{1}{|c|}{Parameter} & 
                \multicolumn{1}{|c|}{Value}     \\
        \hline \hline
        $m_c$                   & $1.4$ (GeV) \\
        $m_b$                   & $4.8$ (GeV) \\
        $\alpha_{em}^{-1}$      & 129           \\
        $\lambda_t$            & 0.04 \\
        $\Gamma_{tot}(B_s)$             & $4.09 \cdot 10^{-13}$ (GeV)   \\
        $f_{B_s}$             & $0.2$ (GeV)  \\   
        $m_{B_s}$             & $5.369$ (GeV) \\
        $m_{t}$             & $175$ (GeV) \\
        $m_{W}$             & $80.26$ (GeV) \\
        $m_{Z}$             & $91.19$ (GeV) \\
        $\Lambda^{(5)}_{QCD}$             & $0.214$ (GeV) \\
        $\alpha_{s}(m_Z)$             & $0.117$  \\
        $\lambda_2$             & $0.12$ $(\mbox{GeV}^2)$ \\
        \hline
        \end{tabular}
        \end{center}
\caption{Values of the input parameters used in the numerical
          calculations unless otherwise specified.}
\label{input}
\end{table}
\section{Constraint analysis}
There is a considerable interest in the constraints of the parameter 
space of the 2HDM, especially in model  II, 
since its Higgs sector coincides with the minimal
supersymmetric extension of the SM one. The free parameters
of the 2HDM are the masses of the charged and neutral Higgs bosons
and the ratio of the v.e.v. of the two Higgs fields, 
denoted by $\tan\beta$. 
In our analysis the neutral Higgs bosons are irrelevant, since they
do not give any contribution to the $b\rightarrow s\gamma\gamma$ process.
Therefore we consider as free parameters the 
mass $m_{H}$ of the charged Higgs boson and $tan\beta$. 
By using existing experimental data, it is possible to find restrictions 
on the parameters $m_{H}$ and $tan\beta$.

The model independent lower bound of the mass of the charged Higgs
$m_{H}\geq 44 \,\,GeV$ comes from the non-observation of charged H
pairs in Z decays \cite{Abreu}.  There are no experimental upper
bounds for $m_{H}$ except $m_{H}\leq 1\,\, TeV$
to satisfy the unitarity condition \cite{Veltman}.
For model II, the constraints have already been studied. Top decays give 
$m_{H}\geq 147\,\,GeV$ for large $tan\beta$ \cite{Abe}.
The lower bound of $tan\beta$ is found to be $0.7$ due to the decay 
$Z\rightarrow b\bar{b}$ \cite{Grant} and in addition for large $tan\beta$ 
the ratio $tan\beta / m_{H}$ is restricted. The current limits are 
$tan\beta / m_{H}\leq 0.38\,\, GeV^{-1}$ \cite{Acci} and 
$tan\beta / m_{H}\leq 0.46\,\, GeV^{-1}$ \cite{Aleph}, 
which come from the experimental
results of branching ratios of the decays $B\rightarrow \tau\bar{\nu}$ 
and $B\rightarrow X\tau\bar{\nu}$.
Recently, the exclusive decay mode $B\rightarrow D\tau\bar{\nu}$ has
been studied \cite{Sonikiers} for model II. It was shown that this decay 
could be used to put an upper bound on $tan\beta/m_{H}$, with the sufficient
data and the reduction of theoretical uncertainities. Under these
conditions, the upper bound was estimated as  
$tan\beta/m_{H} = 0.06\,\, GeV^{-1}$. 

In the present work, we estimate the constraints for the 2HDM parameters
using the result coming from the measurement of the decay $B\rightarrow
X_{s}\gamma$ by the CLEO collaboration \cite{cleo}: 
\begin{eqnarray}
Br (B\rightarrow X_{s}\gamma)=(2.32\pm 0.57\pm 0.35) \cdot 10^{-4}.
\label{branching}
\end{eqnarray}     
To reduce the b-quark mass dependence let us consider the ratio 
\begin{eqnarray}
R&=&\frac{Br (B\rightarrow X_{s}\gamma)}
{Br (B\rightarrow X_{c} e \bar{\nu}_{e})}\nonumber \\
&=& \frac{|V_{ts}^{*} V_{tb}|^{2}}{|V_{cb}|^{2}}\frac{6 \alpha_{em}}
{\pi g(z)} |C_{7}^{eff}|^{2} \, \, ,
\label{R}
\end{eqnarray}
where $g(z)$ is the phase space factor in semileptonic b-decay,
\begin{eqnarray}
g(z)=1-8 z^{2}+8 z^{6}+z^{8}-24 z^{4} ln \,z
\label{gz}
\end{eqnarray}
and $z=m_{c}/m_{b}$.

Now we want to discuss the theoretical uncertainties present in 
the prediction of $R$.
\begin{itemize}

\item The ratio of the CKM matrix elements $\frac{|V_{ts}^{*} V_{tb}|^{2}}{|V_{cb}|^{2}}=0.95\pm 0.04$ 
has  an uncertainty which comes from the CP violating parameter 
$\epsilon_{\kappa}$  \cite{burasmisiak}.

\item The function $g(z)$ has an uncertainty coming from the masses $m_{b}$
and $m_{c}$ via the ratio $z=m_{c}/m_{b}$.
HQET provides a mass relation \cite{Neu}
\begin{eqnarray}
m_{b}-m_{c}=(m_{\bar{B}}-m_{\bar{D}})[1-\frac{\lambda_{1}}{2 m_{B} m_{D}}+
O(\frac{1}{m_{Q}^{3}})] \, \, ,
\label{massdif}
\end{eqnarray}
where $Q=b$ and $c$ , $m_{\bar{B}}$ and $m_{\bar{D}}$ are spin averaged
meson masses,
$m_{\bar{B}}=5.31\,\, GeV$ and $m_{\bar{D}}=1.97\,\, GeV$ \cite{Neu}. 
Here $\lambda_{1}$ is the non-perturbative parameter, which characterizes
the average kinetic energy of the b-quark in $B$ meson and its value 
is obtained by QCD sum rules. 
Using the theoretical estimate for
$\lambda_{1}=-(0.4\pm0.2)\,\, GeV^{2}$ \cite{Ball},
the mass difference and the error quoted are given as
\begin{eqnarray}
m_{b}-m_{c}=(3.40\pm 0.03\pm 0.03)\,\, GeV.
\label{massqdif}
\end{eqnarray}

Here the first error is due to the uncertainty in $\lambda_{1}$
and the second one is from higher order corrections.
We take the central value of b-quark mass as $m_{b}=4.8\,\, GeV$.   
The uncertainty in $m_{b}$ is $\Delta m_{b}=\pm 0.1 \,\, GeV$. Using 
the HQET result, we estimate the uncertainty in $m_{c}$ as 
$\Delta m_{c}=\pm 0.16$ and we get the error in  $g(\bar{z})$
as $\Delta g(\bar{z})=0.096$ and $\Delta g(\bar{z})/g(\bar{z})=
17.8\,\, \%$, where $\bar{z}=m_{c}/m_{b}=0.29$ is the central value.
\end{itemize}

The $Br$ for the semileptonic $B\rightarrow X_{c} e \bar{\nu}_{e}$ is 
\cite{PDG}
\begin{eqnarray}
Br (B\rightarrow X_{c} e \bar{\nu}_{e})=0.103\pm 0.01.
\label{brcenu}
\end{eqnarray}
Both the theoretical uncertainties and the experimental errors, 
as given in eq.~(\ref{branching}) and ~(\ref{brcenu}), result in an 
uncertainty in $C_{7}^{eff}$.
Using 
\begin{eqnarray}
Br_{max}(B\rightarrow X_{s}\gamma)&=&3.24 \cdot 10^{-4} \, \, , \nonumber \\
Br_{min}(B\rightarrow X_{s}\gamma)&=&1.40 \cdot 10^{-4} \, \, ,
\label{Brmaxmin}
\end{eqnarray}
we get a possible range for $|C_{7}^{eff}|$ as
\begin{eqnarray}
0.1930 \leq |C_{7}^{eff}| \leq 0.4049 \, \, . 
\label{C7}
\end{eqnarray}
In the SM and 2HDM model II is $C_{7}^{eff} < 0$, but in possible extensions 
it can be positive. 

Now we present the lower bounds of 
$m_{H}$ for different values of the scale $\mu$ in Table~(\ref{mhlimit}) 
for model II. We restricted $|C_{7}^{eff, 2HDM}(\mu)|$ by using the limits
given in eq.~(\ref{C7}). 
\begin{table}[h]
        \begin{center}
        \begin{tabular}{|c|l|l|l|}
        \hline
        \multicolumn{1}{|c|}{$m_{H min}\,\, [GeV$]}     &
        \multicolumn{1}{|c|}{$\mu\,\, [GeV]$}   \\
        \hline
        \multicolumn{1}{|c|}{$480$}     & 
        \multicolumn{1}{|c|}{$2.5$}\\ 
        \multicolumn{1}{|c|}{$302$}     & 
                \multicolumn{1}{|c|}{$5$ }      \\
        \multicolumn{1}{|c|}{$235$}     & 
        \multicolumn{1}{|c|}{$10$ }     \\
        \multicolumn{1}{|c|}{$158$}     & 
        \multicolumn{1}{|c|}{$m_{W}$ }  \\
        \hline 
        \end{tabular}
        \end{center}
\caption{The lower bounds of the Higgs mass $m_{H}$
for different scales $\mu$ in model II. }
\label{mhlimit}
\end{table}
For model I, a lower bound for the Higgs boson mass is absent. 

\begin{figure}[htb]
\vskip -1.5truein
\centering
\epsfxsize=3.8in
\leavevmode\epsffile{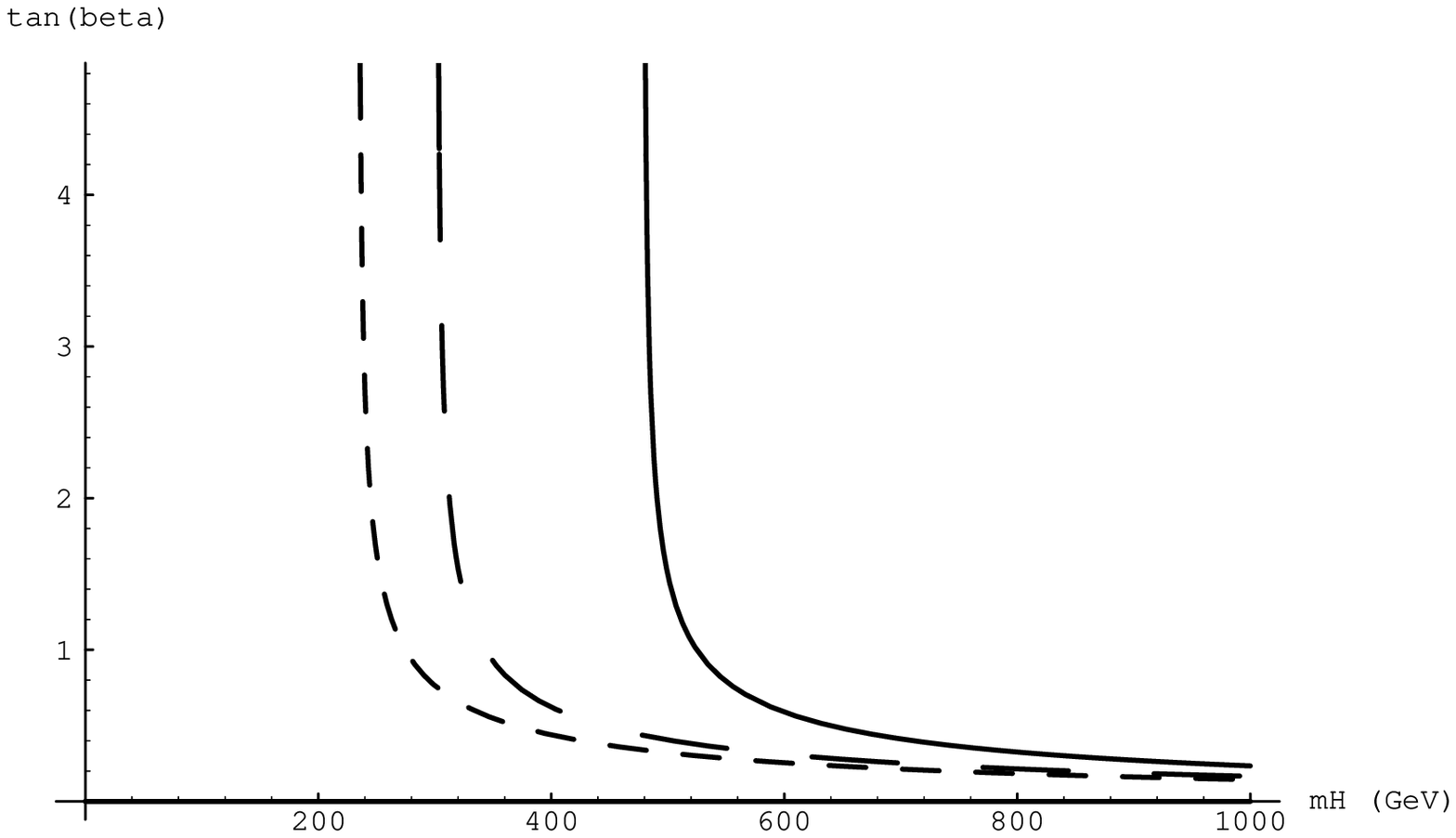}
\vskip -1.5truein
\caption[]{$tan\beta$ as a function of the mass $m_{H}$ 
for fixed $C_{7}^{eff}=-0.4049$ in the model II of the 2HDM.
Here solid curves correspond to the scale $\mu=2.5\,\, GeV$, 
dashed curves to $\mu=5\,\, GeV$ and 
small dashed curves to $\mu=10\,\, GeV$. }
\label{tbetmh}
\end{figure}

The parameter $tan\beta$ has bounds strongly depending on the scale $\mu$
and $m_{H}$. 
In  fig.~(\ref{tbetmh}), we plot the parameter $tan\beta$ with respect 
to $m_{H}$ for 3 different $\mu$ scales  $(2.5,\,\,5,\,\,10) \,\, GeV$ in 
model II,
by fixing  $|C_{7}^{eff, 2HDM}|=0.4049$. 
We see, that the dependence of $tan\beta$ ($m_{H}$) on $m_{H}$ 
($tan\beta$) becomes weak for large values of $m_{H}$ ($tan\beta$) 
and that a decreasing $\mu$ scale causes the allowed
region in the $tan\beta$ - $m_{H}$ plane is to be small. 
It is interesting that 
at $\mu=2.5\,\, GeV$, the solution for $tan\beta$ - $m_{H}$ 
exists only in
the region $0.4047 \leq |C_{7}^{eff, 2HDM}|\leq 0.4049$. 
Therefore the solid curve in
Fig~(\ref{tbetmh}) is almost the allowed region for 
$(tan\beta,m_{H})$.
For $\mu=2.5\,\,GeV$, we get an empirical expression for the 
restricted region of the parameter set $(tan\beta,m_{H})$, 
\begin{eqnarray}
tan\beta= c_{1}+\frac{c_{2}}{\sqrt{m_{H}-m_{p}}} \, \, ,
\end{eqnarray}
with $c_{1}=-0.067$, $c_{2}=6.9\,\, GeV^{1/2}$ 
and $m_{p}=m_{H\, min}-\epsilon$. Here $\epsilon$ is a positive small number 
$(\epsilon \ll 1 \, )$ and $m_{H\, min}=480\,\, GeV$.

In the following analysis we restrict the coefficient 
$|C_{7}^{eff, 2HDM}|$ in the
given region and study the resulting $m_{H}$, $tan\beta$ and scale $\mu$ 
dependencies of the ratio $|A^{+}|^{2}/|A^{-}|^{2}$ and the $Br$
for the decay $B_{s}\rightarrow \gamma\gamma$.

\section{Discussion}
In the rest frame of the $B_s$ meson, the $CP=-1$ amplitude $A^{-}$
is proportional to the perpendicular spin polarization
$\vec{\epsilon_1}\times\vec{\epsilon_2}$, and the $CP=1$ amplitude $A^{+}$ is
proportional to the parallel spin polarization
$\vec{\epsilon_1}.\vec{\epsilon_2}$. 
The ratio $|A^{+}|^2/|A^{-}|^2$ is informative to search for CP violating 
effects in $B_{s}\rightarrow\gamma\gamma$ decays and it has been studied 
before in the literature in the framework of the 2HDM
without QCD corrections \cite{aliev}.
In our analysis we use three sets of parameters 
$(m_b,\bar{\Lambda}_s)$ given in (Table~(\ref{mums})), which
model the bound state \cite{Gud}. However, 
we do not present the figures for the first two. 
Here we analyze the LLog $\mu$, and 2HDM parameters ($m_{H}$, $tan\beta$)
dependence of the ratio $|A^{+}|^2/|A^{-}|^2$ 
and  present the  results in a series of graphs (fig. 2-7).

\begin{figure}[htb]
\vskip -1.5truein
\centering
\epsfxsize=3.8in
\leavevmode\epsffile{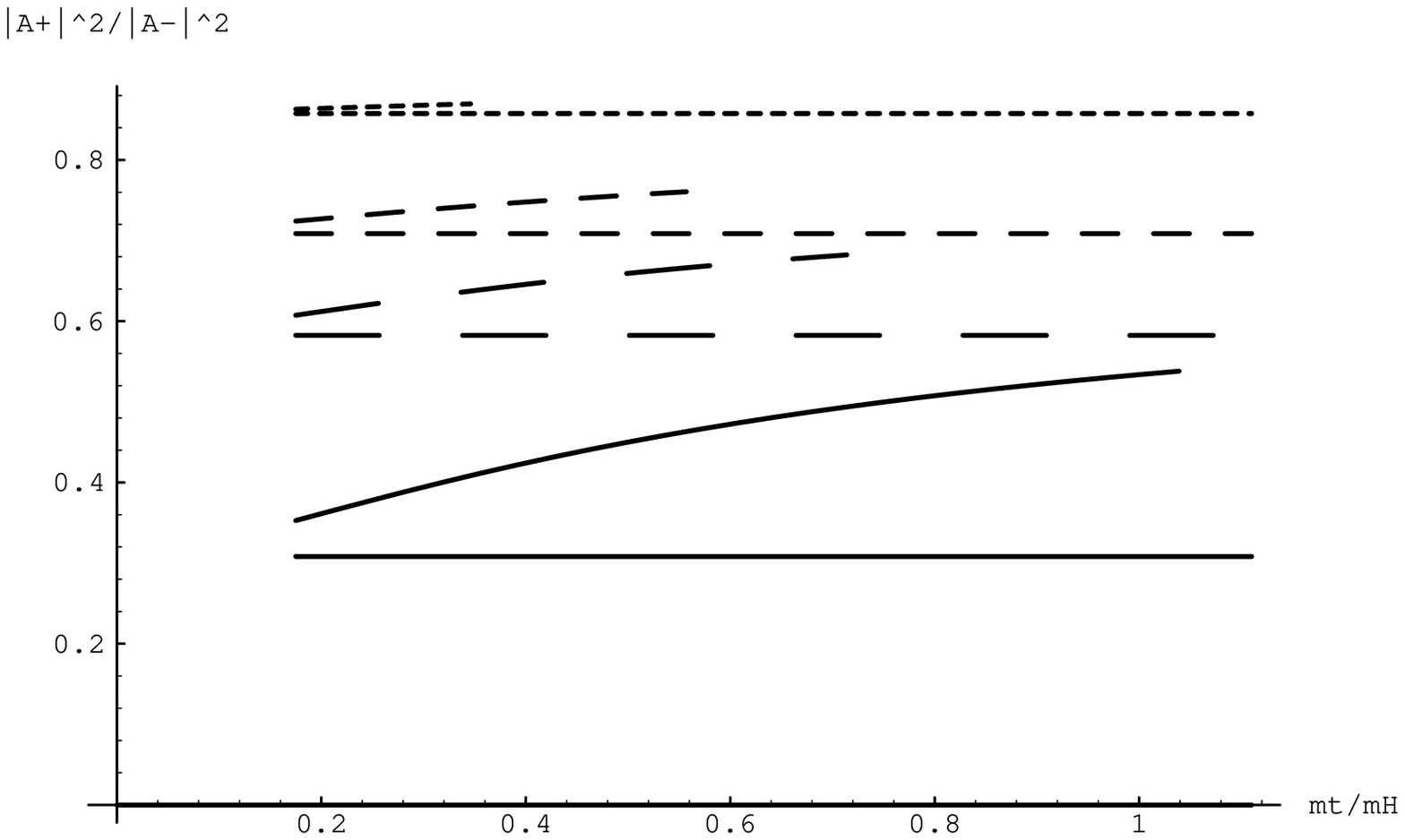}
\vskip -1.5truein
\caption[]{$m_{t}/m_{H}$ dependence of the ratio $|A^{+}|^2/|A^{-}|^2$
for set 3 and $tan\beta=2$. 
Here, solid lines (curves) correspond to the 
SM  (model II 2HDM) at $\mu=m_{W}$, long dashed lines (curves) to SM 
(model II 2HDM) 
at $\mu=10\,\, GeV$, medium dashed lines (curves) to SM (model II 2HDM) 
at $\mu=5\,\, GeV$  and small dashed lines (curves) to SM (model II 2HDM) 
at $\mu=2.5\,\, GeV$.}
\label{ksi05p3}
\end{figure}

\begin{figure}[htb]
\vskip -1.5truein
\centering
\epsfxsize=3.8in
\leavevmode\epsffile{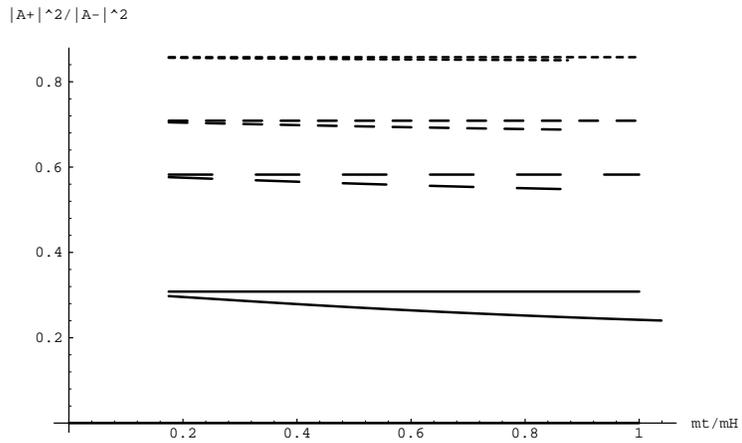}
\vskip -1.5truein
\caption[]{Same as fig~2, but in model I.}
\label{ksi05Ip3}
\end{figure}

In fig.~(\ref{ksi05p3}) and ~(\ref{ksi05Ip3}) 
we plot the dependence of $|A^{+}|^{2}/|A^{-}|^{2}$ 
on $m_{t}/m_{H}$ for fixed $tan\beta=2$  and four  
different $\mu$ scales, $(m_{W},\,10,\,5,\,2.5)\,\, GeV$
in model II and model I, respectively. 
Decreasing the scale $\mu$ weakens the dependence of
the ratio $|A^{+}|^{2}/|A^{-}|^{2}$ on $m_{H}$ and the contribution of 
the charged Higgs bosons to $|A^{+}|^{2}/|A^{-}|^{2}$ gets small. 
The lower limit of the Higgs mass is 
sensitive to the scale $\mu$ and it increases with decreasing 
$\mu$ in  model II, (Table~(\ref{mhlimit})). 
However, the Higgs mass has no lower bound in model I.

\begin{figure}[htb]
\centering
\vskip -1.5truein
\epsfxsize=3.8in
\leavevmode\epsffile{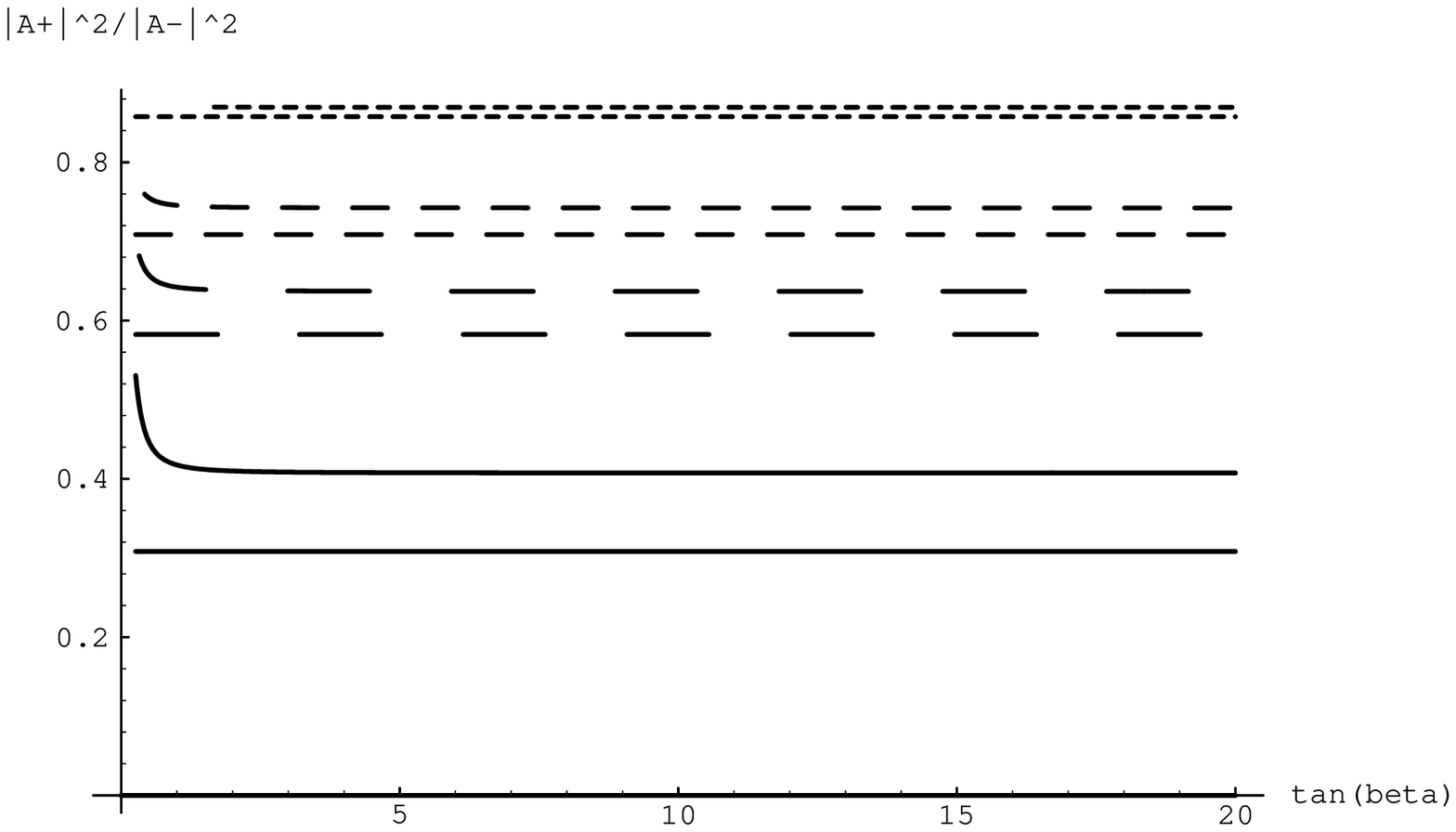}
\vskip -1.5truein
\caption[]{$tan\beta$ dependence of the ratio $|A^{+}|^2/|A^{-}|^2$
for set 3 for $m_{H}=500\,\, GeV$. 
Here, solid lines (curves) correspond to
the SM (model II 2HDM) at $\mu=m_{W}$, long dashed lines (curves) to SM 
(model II 2HDM) 
at $\mu=10\,\, GeV$, medium dashed lines (curves) to SM (model II 2HDM) 
at $\mu=5\,\, GeV$ and small dashed lines (curves) to SM (model II 2HDM) 
at $\mu=2.5\,\, GeV$.}
\label{yt500p3}
\end{figure}

\begin{figure}[htb]
\vskip -1.5truein
\centering
\epsfxsize=3.8in
\leavevmode\epsffile{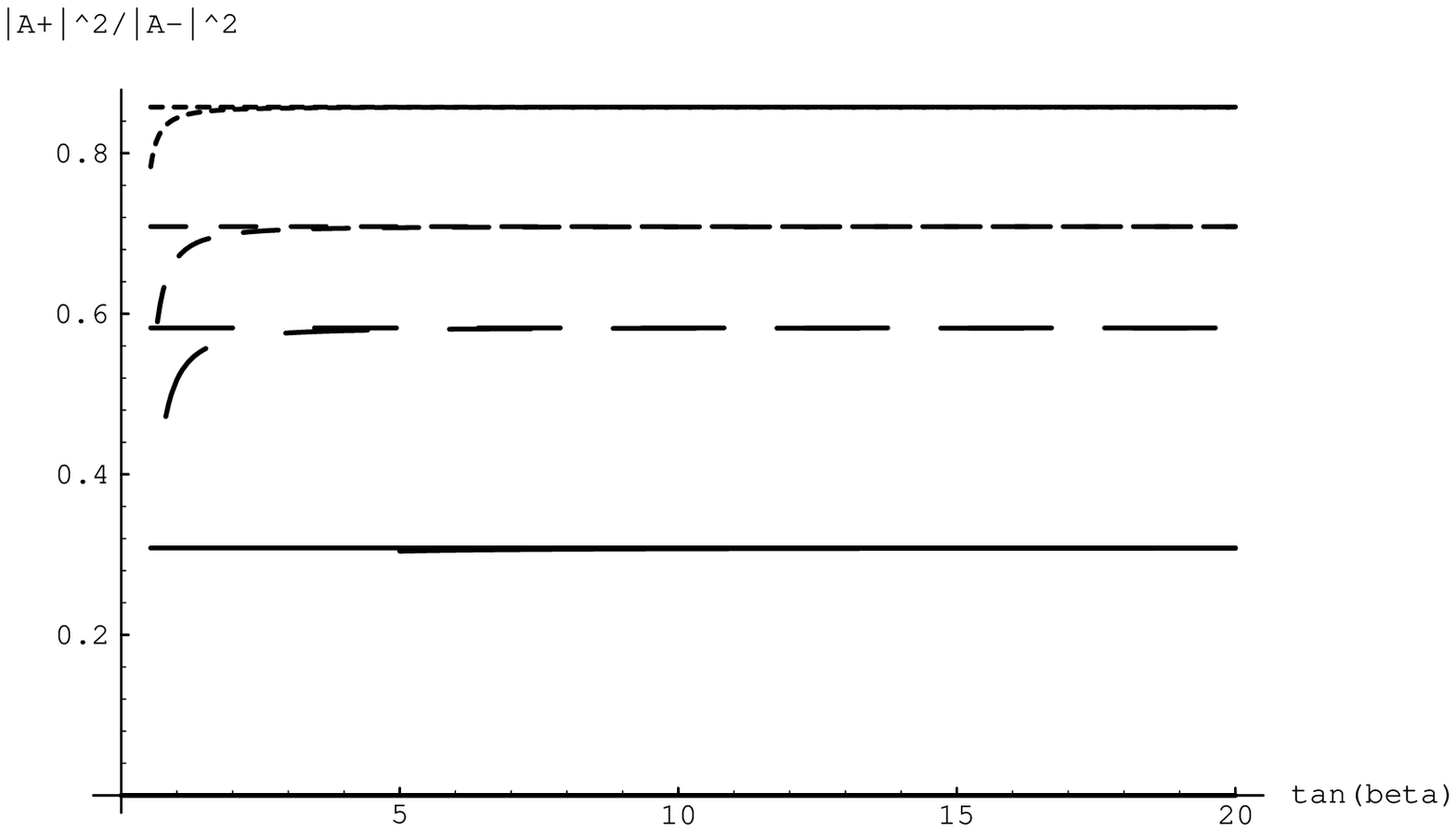}
\vskip -1.5truein
\caption[]{Same as fig~4, but in model I 2HDM.}
\label{yt500Ip3}
\end{figure}

Fig.~(\ref{yt500p3}) and ~(\ref{yt500Ip3}) show the dependence 
of $|A^{+}|^{2}/|A^{-}|^{2}$ on $tan\beta$ for 
fixed $m_{H}=500\,\,GeV$. This ratio is sensitive  only to
small $tan\beta$ values.  The $\mu$ scale 
regulates the lower limit of $tan\beta$ for 
model I and model II in an opposite way.  
Further, the effect of the charged Higgs contribution becomes weak
for large $tan\beta$ values. 

\begin{figure}[htb]
\vskip -1.5truein
\centering
\epsfxsize=3.8in
\leavevmode\epsffile{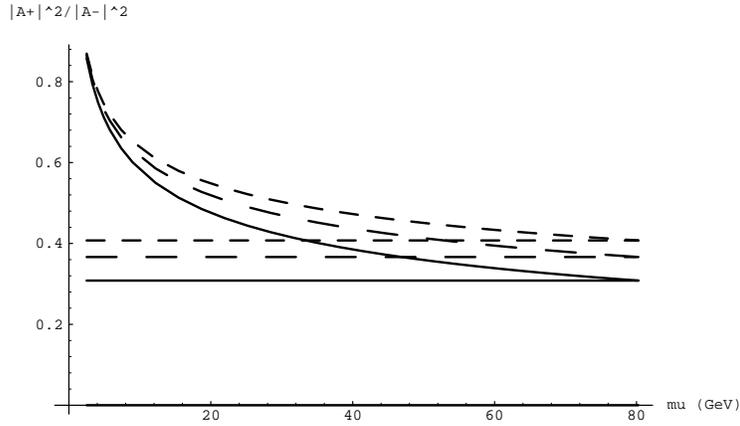}
\vskip -1.5truein
\caption[]{The scale dependence of the ratio $|A^{+}|^2/|A^{-}|^2$
for set 3 in the SM and for 2 values of $m_{H}=500, \,\, 800\,\,GeV$ 
with $tan\beta=10$. 
Here, solid lines (curves) correspond to the SM
at $\mu=m_{W}$ (at arbitrary $\mu$ scale) , long dashed lines (curves) to 
model II with $m_{H}=800\,\, GeV$ 
at $\mu=m_{W}$  (at arbitrary $\mu$ scale),
and small dashed lines (curves) model II with $m_{H}=500\,\, GeV$ 
at $\mu=m_{W}$ (at arbitrary $\mu$ scale).}
\label{mu10p3}
\end{figure}

\begin{figure}[htb]
\vskip -1.5truein
\centering
\epsfxsize=3.8in
\leavevmode\epsffile{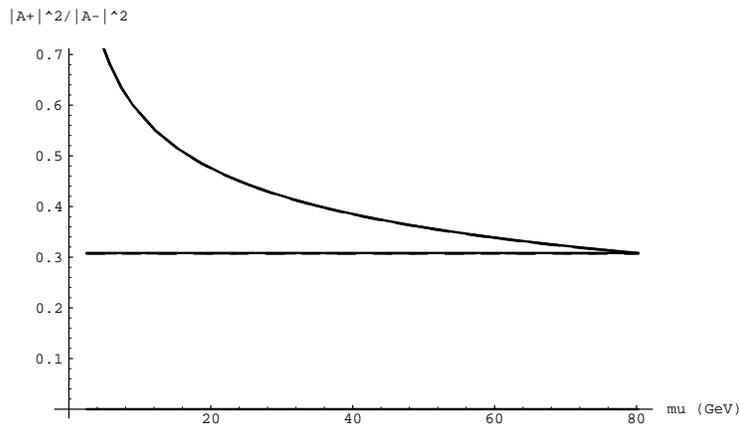}
\vskip -1.5truein
\caption[]{Same as fig~6, but for model I. 
All curves coincide within errors.}
\label{mu10Ip3}
\end{figure}

In fig.~(\ref{mu10p3}) and ~(\ref{mu10Ip3}) we present the $\mu$ 
scale dependence of
$|A^{+}|^{2}/|A^{-}|^{2}$ for the SM and 2HDM with  
$tan\beta=10$ and two different mass values $m_{H}=500\,\,GeV,\,\,
800\,\, GeV$ for model II and model I, respectively.
We find, that for model II the smaller the value of $m_{H}$, 
the less dependent is the ratio on $\mu$. Model I does not allow us to
discriminate between the SM or different values of $m_{H}$ used as 
expected (see e.g. fig.~(5)). 
\begin{table}[h]
        \begin{center}
        \begin{tabular}{|c|l|l|l|}
        \hline
        \multicolumn{1}{|c|}{$set\,\, 1$}       & 
        \multicolumn{1}{|c|}{$set\,\, 2$}       & 
        \multicolumn{1}{|c|}{$set\,\, 3$}       \\ 
         \hline 
                \multicolumn{1}{|c|}{$\bar{\Lambda}_s=370$ MeV} 
& 
                \multicolumn{1}{|c|}{$\bar{\Lambda}_s=480$ MeV}
& 
                \multicolumn{1}{|c|}{$\bar{\Lambda}_s=590$ MeV}\\
\multicolumn{1}{|c|}  {$m_b=5.03$ GeV} & $m_b=4.91$ GeV & $m_b=4.79$ GeV \\
        \hline 

        \end{tabular}
        \end{center}
\caption{The parameter sets of the bound 
state model, $(m_b,\bar{\Lambda}_s)$. }
\label{mums}
\end{table}

The lowest order result of $|A^{+}|^{2}/|A^{-}|^{2}$ 
in $\alpha_{s}$ is obtained
by setting  $\mu=m_{W}$ and it is 
$0.30$ in the SM for set 3.
It reaches $0.85$ at $\mu=2.5\,\, GeV$. 
Varying $\mu$ in the range 
$2.5 \, \,{\mbox{GeV}} \leq \mu \leq 10.0 \, \,{\mbox{GeV}}$,
$|A^{+}|^{2}/|A^{-}|^{2}$ is changing in the range 
$0.60 \, \, \leq |A^{+}|^{2}/|A^{-}|^{2} \leq 0.85$, resulting in
an uncertainty of $\frac{\triangle (|A^{+}|^2/|A^{-}|^2)}{(
|A^{+}|^2/|A^{-}|^2) (\mu=5\,\,GeV)}
\approx \pm 35 \%$ in the SM.
Now we give an example to compare the 
dependence of the ratio $|A^{+}|^2/|A^{-}|^2$ on the 
scale $\mu$ and the 2HDM parameters:
In model II, 
for $m_{H}=500\,\, GeV$ and $tan\beta\geq 2$,
the lowest order result of the ratio $|A^{+}|^2/|A^{-}|^2$ is 
$0.40$ and it enhances up to $0.50$ with decreasing $tan\beta$. 
However, at $\mu=2.5 \,\, GeV$ the ratio reaches $0.86$ 
and the uncertainty due to the extended
Higgs sector is weaker than the one due to the scale $\mu$.
In model I, the behaviour is the same.

This shows, that the ratio $|A^{+}|^{2}/|A^{-}|^{2}$ 
is quite sensitive to QCD corrections
and this strong  $\mu$ dependence makes the  
analysis of the 2HDM parameters $m_{H}$ and $tan\beta$ for the given
experimental value of the ratio $|A^{+}|^{2}/|A^{-}|^{2}$ difficult.
However, we believe, that
the strong $\mu$ dependence will be reduced 
with the addition the of next to leading order (NLO) calculation, 
and the analysis on the parameters will be more reliable.
Note, that a similar analysis for the decay 
$b\rightarrow s\gamma$ is given in \cite{misiak}, \cite{burasmisiak}. 

In addition, there is another uncertainty due to the different parameter
sets (Table~(\ref{mums})). For set 2 $|A^{+}|^{2}/|A^{-}|^{2}=
0.36$ and for set 1 $|A^{+}|^{2}/|A^{-}|^{2}=0.40$ in the SM and 
in the lowest order of $\alpha_s$.
It follows, that the larger 
$m_{b}$ (smaller $\bar{\Lambda}_{s}$), the larger the ratio.
We further see that having increased $m_{b}$, the ratio 
$|A^{+}|^{2}/|A^{-}|^{2}$ becomes less sensitive to the scale $\mu$. 
This ratio essentially changes when QCD corrections
are taken into account. In the lowest order of $\alpha_s$,  
$A^{+}$ and $A^{-}$ depend both on the one
particle reducible part (IPR), proportional to $C_7^{eff}=C_{7}$ 
and in addition $A^-$ contains the
one particle irreducible part (IPI), proportional to 
$C_2$, see eq.~(\ref{Amplitudes}). If we include
QCD corrections to the considered ratio, the contribution of  $C_7^{eff}$
dominates over the IPI sector and the value of the ratio increases.
This means, that the values of $A^+$ and $A^-$ come close to each other and 
it can be explained as a cancellation of the IPI sector.

Now we continue to analyze the $Br$ displayed 
in a series of figures 8-13.
In fig. 8-11, we present the $\frac{m_{t}}{m_{H}}$ and 
$tan\beta$ dependencies of the $Br$. Decreasing $m_{H}$,
the $Br$ increases in model II, however, the behaviour is opposite in
model I. On the other hand the $Br$ is sensitive to small $tan\beta$.
For large values of $tan\beta$ , 
the dependence of the $Br$ on $tan\beta$ becomes weak in model II.
In model I, the 2HDM result almost coincides with the SM one
since the charged Higgs contribution is proportional
to $1/(tan\beta )^2$. 
Similar to the case of $|A^{+}|^2/|A^{-}|^2$, the $Br$ 
is strongly
dependent on the scale $\mu$, see figs~(12-13)). It is enhanced for small
values of $\mu$. For parameter set 3, the lowest order result is
$3.6 \cdot 10^{-7}$ in the SM. It increases up to $6.8 \cdot 10^{-7}$ at 
$\mu=2.5\,\,GeV$. Varying $\mu$ in the range 
$2.5 \, \,{\mbox{GeV}} \leq \mu \leq 10.0 \, \,{\mbox{GeV}}$, 
the $Br$ changes between
$5.0 \cdot 10^{-7} \leq Br \leq 6.9 \cdot 10^{-7}$, and this  gives
an uncertainty 
$\triangle{Br}/Br (\mu=5\,\,{\mbox{GeV}})\approx \pm 30 \%$ 
for the SM. 
For set $1\,\, (2)$, the $Br$ increases up to $1.7 \cdot 10^{-6}$
($1.0 \cdot 10^{-6}$) and the uncertainty is also increases, 
as due to the scale $\mu$ dependence, namely $39\, (35) \,\%$. 
This behaviour of the $Br$ results mainly from 
the $1/\bar{\Lambda}_{s}$  dependence in amplitudes.

With the addition of extra Higgs contribution, the uncertainty due to 2HDM
parameters $m_{H}$ and $tan\beta$ in the $Br$ becomes large
like the one coming from the scale $\mu$.
Now we will give an example to see the effect of the 2HDM parameters
on the $Br$ by choosing set 3. 
In model II
for $m_{H}=500\,\, GeV$,
the lowest order result of the $Br$  
is $5.8 \cdot 10^{-7}$ for $tan\beta \geq 2$ (see fig.~10) and it reaches 
$1.1 \cdot 10^{-6}$ for smaller
$tan\beta$. For comparison, the value in
the SM is $3.6 \cdot 10^{-7}$. 
At $\mu=2.5 \,\, GeV$ the $Br$ reaches $6.9 \cdot 10^{-7}\,\,
(9.0 \cdot 10^{-7})$ 
in the SM (2HDM).. 
This shows, that the $Br$ is also sensitive to the extra Higgs
contribution. For $m_{H}=500\,\, GeV$ and $\mu =2.5\,\, GeV$, 
the $Br$ in the 2HDM model II is enhanced
$\sim 30 \%$ compared to the SM.
In model I, there is a suppression due to the extra Higgs contribution 
compared to the SM (see fig.~(11)), however, 
the $Br$ is still sensitive to the scale $\mu$ (see fig.~(13))
and the 2HDM parameters.

We complete this section by
taking the $O_{7}$ type long distance effects ($LD_{O_{7}}$)
for both the ratio $|A^{+}|^2/|A^{-}|^2$ and the $Br$ into account.
The $LD_{O_{7}}$ contribution to the CP-odd $A^{-}$ and CP-even $A^{+}$ 
amplitudes has recently been calculated with the help of the 
Vector Meson
Dominance model (VMD) \cite{Gud} and it was shown, that the influence on the
amplitudes is destructive. With the addition of the $LD_{O_{7}}$ effects,
the amplitudes entering $|A^{+}|^2/|A^{-}|^2$ and the $Br$ are now given as
\begin{eqnarray}
A^{+}&=& A^{+}_{SD}+A^{+}_{LD_{O_{7}}}\nonumber \,\, , \\
A^{-}&=& A^{-}_{SD}+A^{-}_{LD_{O_{7}}} \,\, , 
\label{LDamp}
\end{eqnarray}
where $A^{\pm}_{SD}$ are the short distance amplitudes we took into account
in the previous sections (eq.~10). The $LD_{O_{7}}$ amplitudes 
$A^{\pm}_{LD_{O_{7}}}$ are defined as \cite{Gud} 
\begin{eqnarray}
A^{+}_{LD_{O_7}}&=&
-\sqrt{2} \frac{\alpha_{em} G_F}{\pi} \bar{F_1}(0) f_{\phi}(0) 
\lambda_t \frac{m_b (m_{B_s}^2-m_{\phi}^2)}{3 m_{\phi} m_{B_{s}}^{2}} 
C_{7}^{eff}(\mu)  \, \, , \nonumber\\
A^{-}_{LD_{O_7}}&=&
\sqrt{2} \frac{\alpha_{em} G_F}{\pi} \bar{F_1}(0) f_{\phi}(0) 
\lambda_t \frac{m_b}{3 m_{\phi}} C_{7}^{eff}(\mu) \, \, ,
\end{eqnarray}
where $f_{\phi}(0)=0.18\,\, GeV$ is the decay constant of $\phi$ meson
at zero momentum, 
$\bar{F}_{1}(0)$ is the extrapolated $B_{s}\rightarrow \phi$ form factor
(for details see \cite{Gud}).

In fig. (~\ref{ksi01p3LD} -~\ref{brmu10p3LD} )
we present the $m_{H}$, $tan\beta$ and $\mu$ dependencies of
the ratio $|A^{+}|^2/|A^{-}|^2$ and the $Br$  with the addition of 
$LD_{O_{7}}$ effects for set 3. 
Here we use $\bar{F}_{1}(0)=0.16$ \cite{Gud}. 
It can be shown, that the value of the ratio $|A^{+}|^2/|A^{-}|^2$ decreases
with the addition of $LD_{O_{7}}$ effects. However, while the scale $\mu$
is decreasing, the effect of the $LD_{O_{7}}$ contribution 
on the ratio $|A^{+}|^2/|A^{-}|^2$ is also decreasing, see fig.~4 and ~15.  
On the other hand, the uncertainty resulting from varying $\mu$ between
$2.5 \, \,{\mbox{GeV}} \leq \mu \leq 10.0 \, \,{\mbox{GeV}}$ has increased
compared to the case without the inclusion of the $LD_{O_{7}}$ amplitudes:
\begin{eqnarray}
\frac{\triangle (|A^{+}|^2/|A^{-}|^2)}{
(|A^{+}|^2/|A^{-}|^2) (\mu=5\,\,{\mbox{GeV}})}
\approx \pm 40 \% \nonumber \, \, . 
\label{LDCPun}
\end{eqnarray}
The $Br$ decreases with the addition of $LD_{O_{7}}$ effects,
since the effect is destructive.  
The $\mu$ scale uncertainty of the $Br$ is smaller compared to the case 
where no LD effect is included and we get for the range 
$2.5 \, \,{\mbox{GeV}} \leq \mu \leq 10.0 \, \,{\mbox{GeV}}$ in the SM
\begin{eqnarray}
\frac{\triangle Br}{
Br(\mu=5\,\,{\mbox{GeV}})}
\approx \pm 27 \%  \, \, .
\end{eqnarray}
The present experimental limit on the decay $B_{s}\rightarrow \gamma\gamma$ is
\cite{Acciarri}
\begin{eqnarray}
Br(B_{s}\rightarrow \gamma\gamma)\leq 1.48 \cdot 10^{-4} \, \, ,
\end{eqnarray} 
which is far from the theoretical results.
By varying the parameters 
$\mu, m_{H}, tan\beta, (m_{b}, \bar{\Lambda}_{s}$), 
it is possible to 
enhance the $Br$ up to $2.1 \, (2.5) \cdot 10^{-6}$ in model II
for $m_H=480$ GeV and large $tan\beta$, 
where the possible maximal value in the SM is $1.4 \, (1.7) \cdot 10^{-6}$,
both at $\mu=2.5$ GeV and for set 1.
The numbers in parentheses correspond to the case where no $LD_{O7}$ effects
are taken into account.

LLog calculations show that the $Br$  strongly depends on the scale
$\mu$. This strong dependence will disappear with the addition of 
NLO QCD corrections. From NLO $b\rightarrow s\gamma$ decay, the choice 
of $\mu=m_{b}/2$ in the LLog expression reproduces effectively the NLO 
result, so one suggest that this may work also for the 
$b\rightarrow s\gamma\gamma$ decay.
An additional theoretical uncertainty arises from the poor knowledge 
of the $B_s$ bound state effects.

We find that 
the $Br$ increases with the addition of the extra Higgs contribution and
even at the scale $\mu=2.5\,\, GeV$ this value is $\sim 2$ orders of 
magnitude smaller than the present experimental upper bound.
The other possibility for an enhancement of the $Br$ is the extension 
of the Higgs sector. 
This forces us to think of further models like MSSM,...etc. and 
FCNC $B_{s}\rightarrow \gamma\gamma$ decay will be an efficient tool to
search for new physics beyond the SM.

\bigskip
\noindent
{\Large \bf Acknowledgements}

One of the authors (T. M. A) would like to thank International Centre for
Theoretical Physics (ICTP) for the warm hospitality.
He also acknowleges Prof.S.Randjbar-Daemi for his interest and support.

\newpage
\newpage

\begin{figure}[htb]
\vskip -1.5truein
\centering
\epsfxsize=3.8in
\leavevmode\epsffile{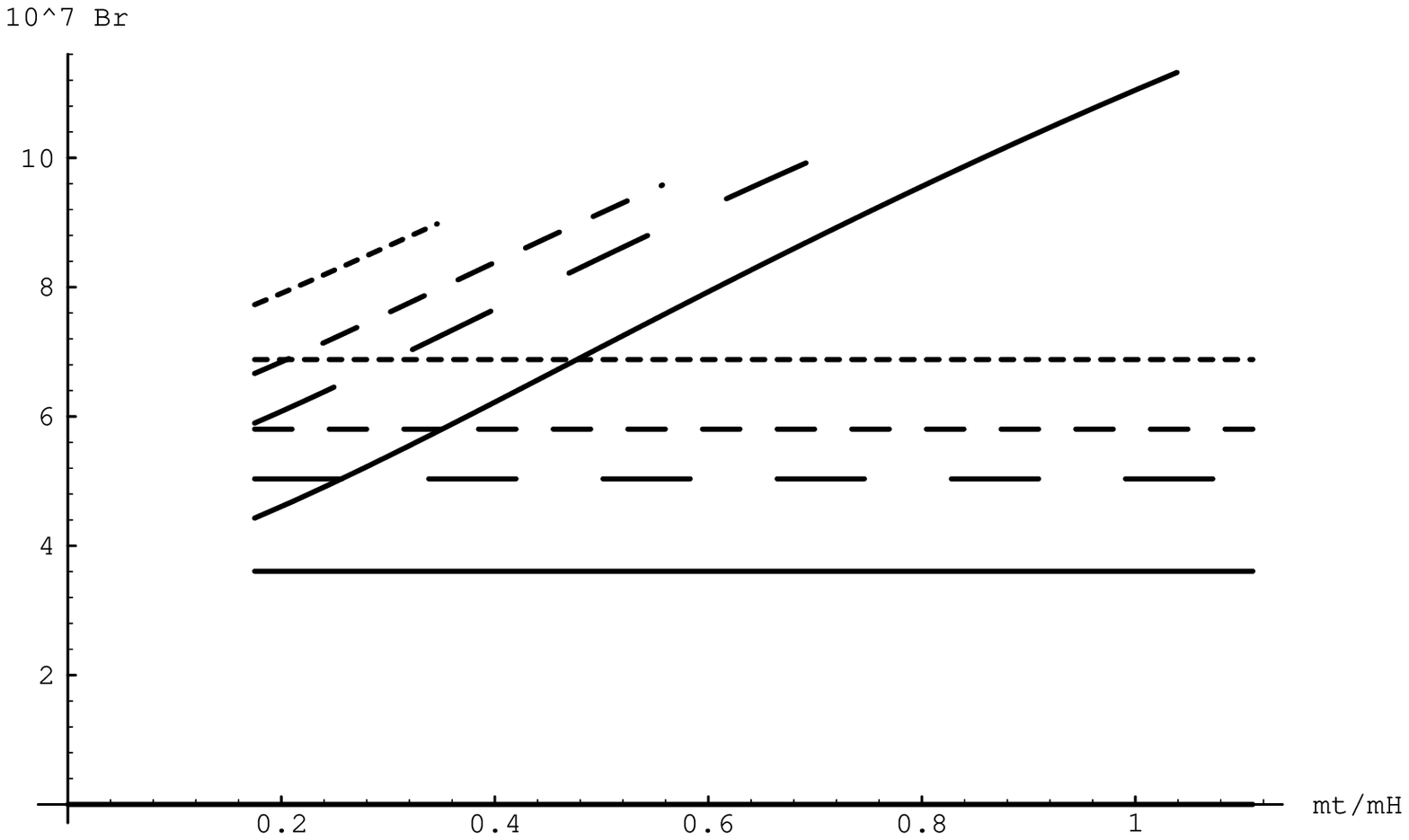}
\vskip -1.5truein
\caption[]{$m_{t}/m_{H}$ dependence of the branching ratio $Br$ 
for set 3 with $tan\beta=2$. 
Here, solid lines (curves) correspond to the 
SM (model II 2HDM) at $\mu=m_{W}$, long dashed lines (curves) to the 
SM (model II 2HDM) 
at $\mu=10\,\, GeV$, medium dashed lines (curves) to the SM (model II 2HDM) 
at $\mu=5\,\, GeV$ and small dashed lines (curves) to the SM (model II 2HDM) 
at $\mu=2.5\,\, GeV$.}
\label{brksi05p3}
\end{figure}

\begin{figure}[htb]
\vskip -1.5truein
\centering
\epsfxsize=3.8in
\leavevmode\epsffile{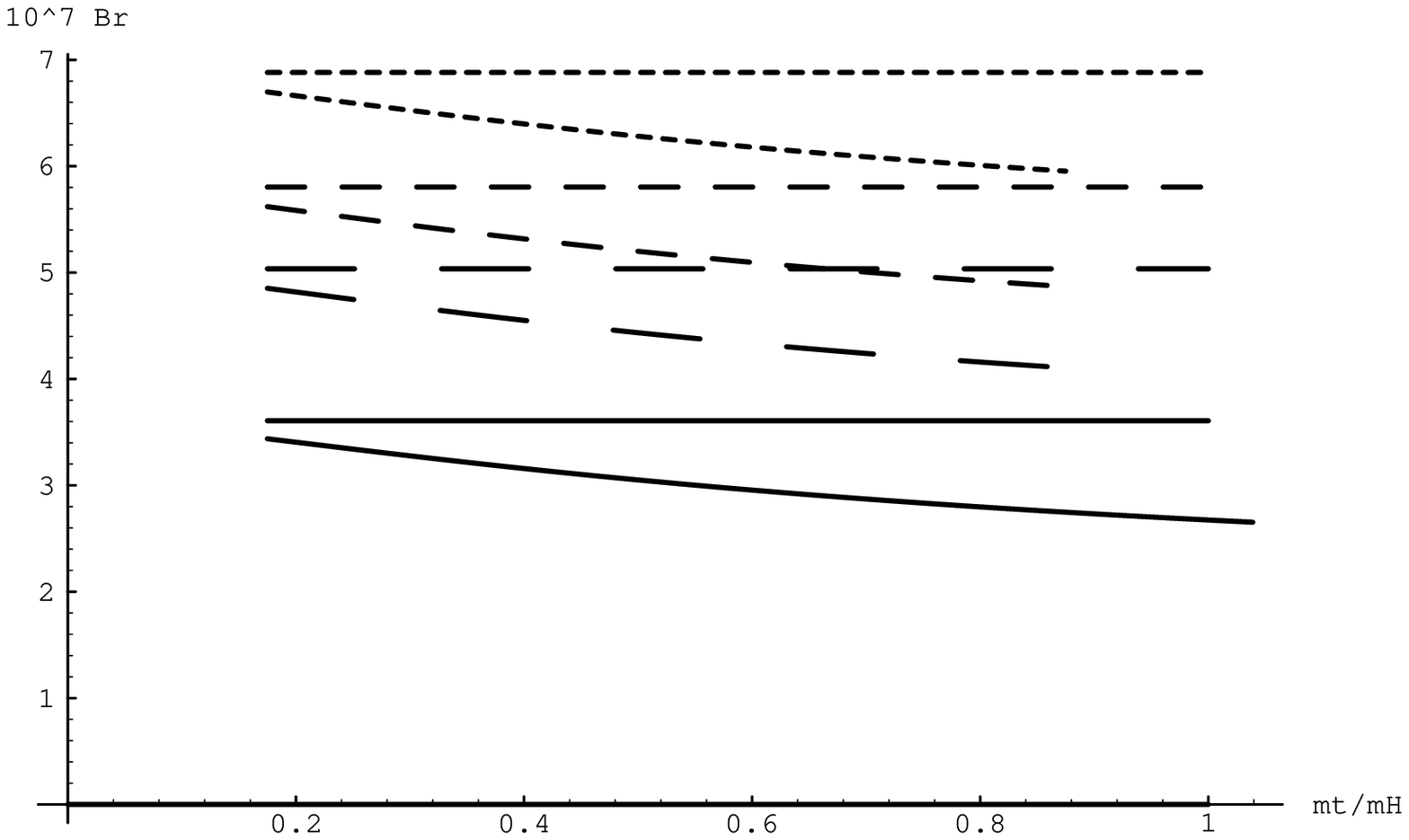}
\vskip -1.5truein
\caption[]{Same as fig~8, but in model I 2HDM.}
\label{brksi05Ip3}
\end{figure}

\begin{figure}[htb]
\vskip -1.5truein
\centering
\epsfxsize=3.8in
\leavevmode\epsffile{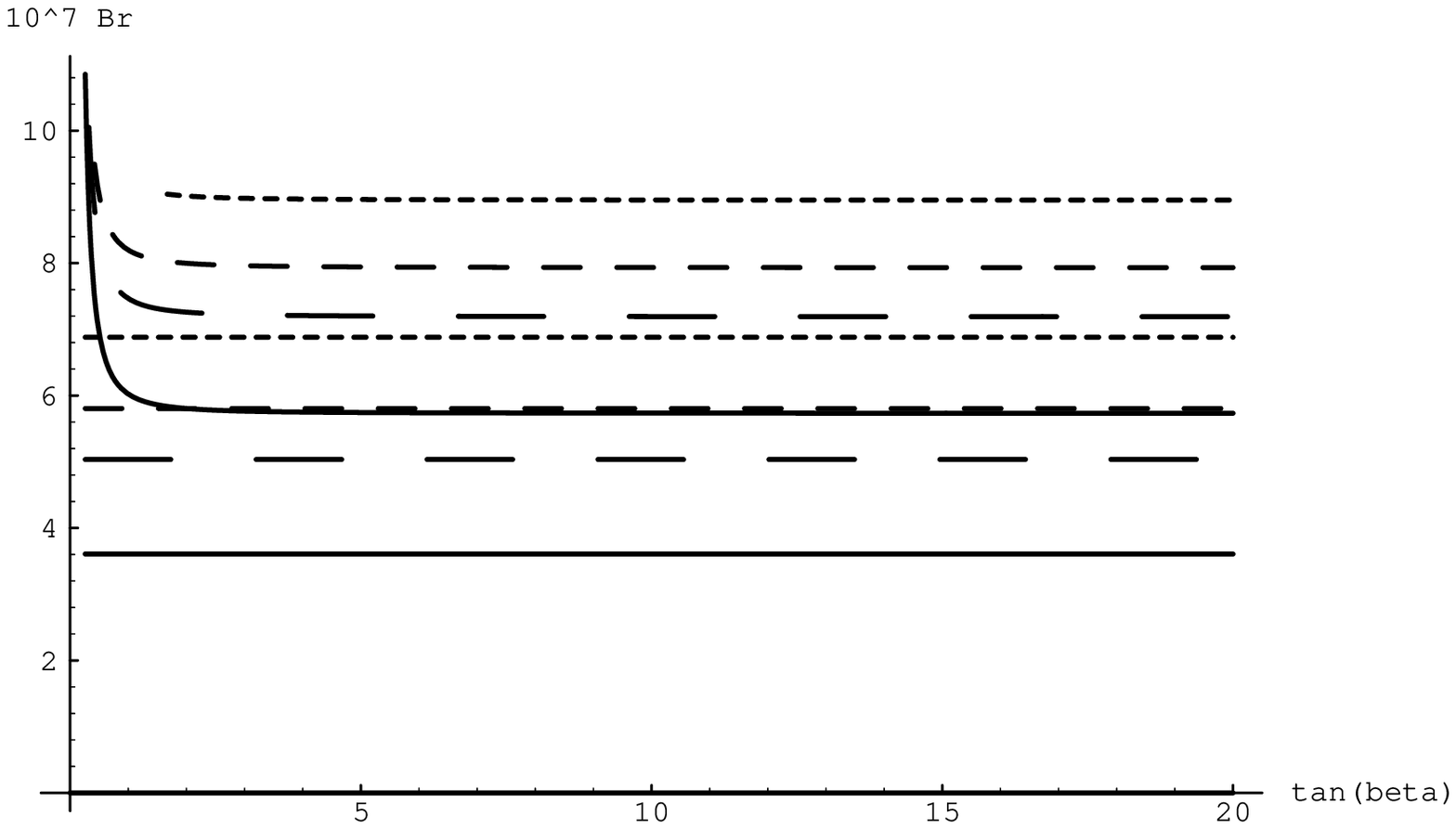}
\vskip -1.5truein
\caption[]{$tan\beta$ dependence of the $Br$ 
for set 3 with $m_{H}=500\,\, GeV$. 
Here, solid lines (curves) correspond to the 
SM  (model II 2HDM) at $\mu=m_{W}$, long dashed lines (curves) to the 
SM (model II 2HDM) 
at $\mu=10\,\, GeV$, medium dashed lines (curves) to the SM (model II 2HDM) 
at $\mu=5\,\, GeV$ and small dashed lines (curves) to the SM (model II 2HDM) 
at $\mu=2.5\,\, GeV$.}
\label{bryt500p3}
\end{figure}

\begin{figure}[htb]
\vskip -1.5truein
\centering
\epsfxsize=3.8in
\leavevmode\epsffile{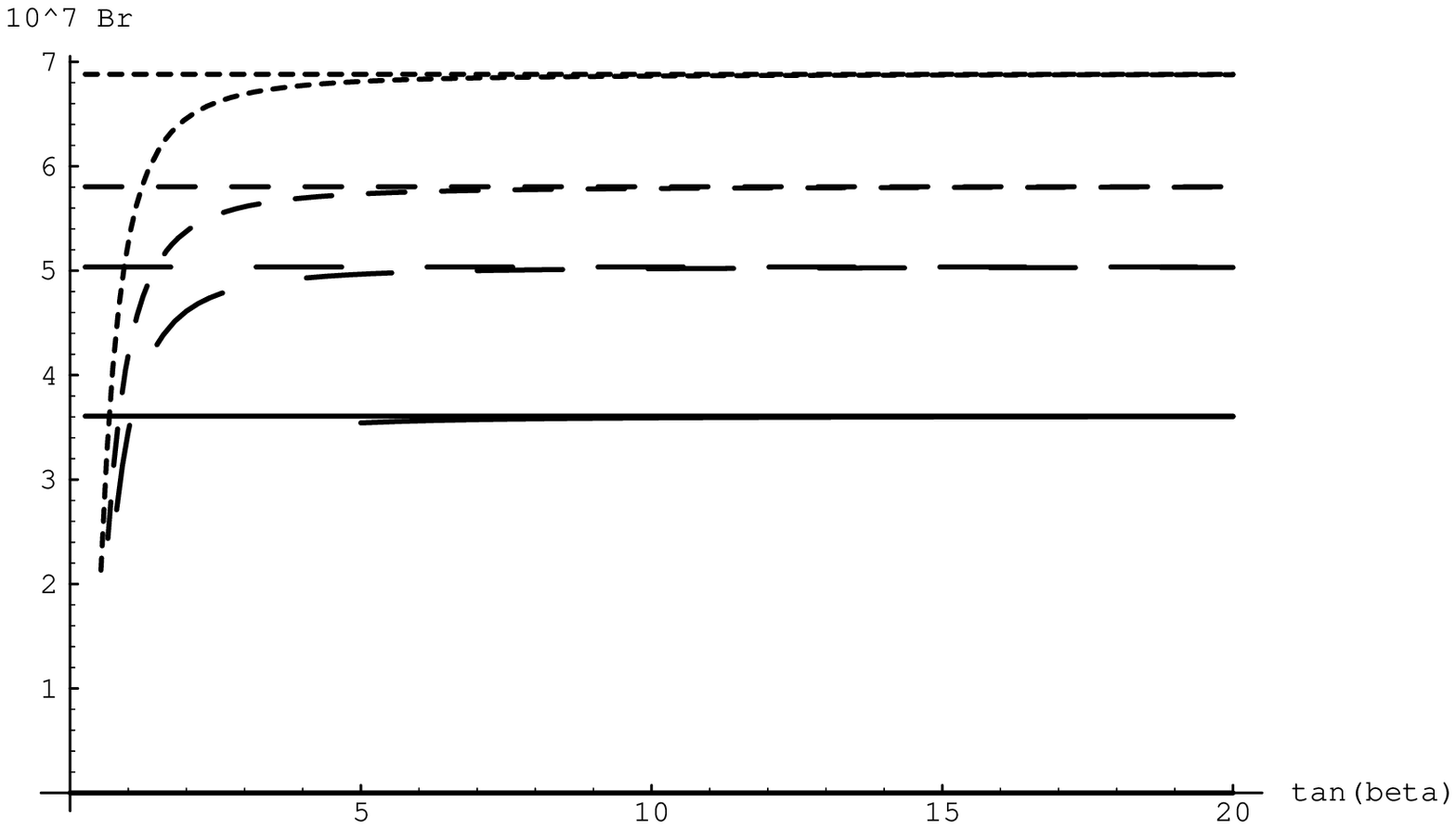}
\vskip -1.5truein
\caption[]{Same as fig~10, but in model I 2HDM.}
\label{bryt800Ip3}
\end{figure}

\begin{figure}[htb]
\vskip -1.5truein
\centering
\epsfxsize=3.8in
\leavevmode\epsffile{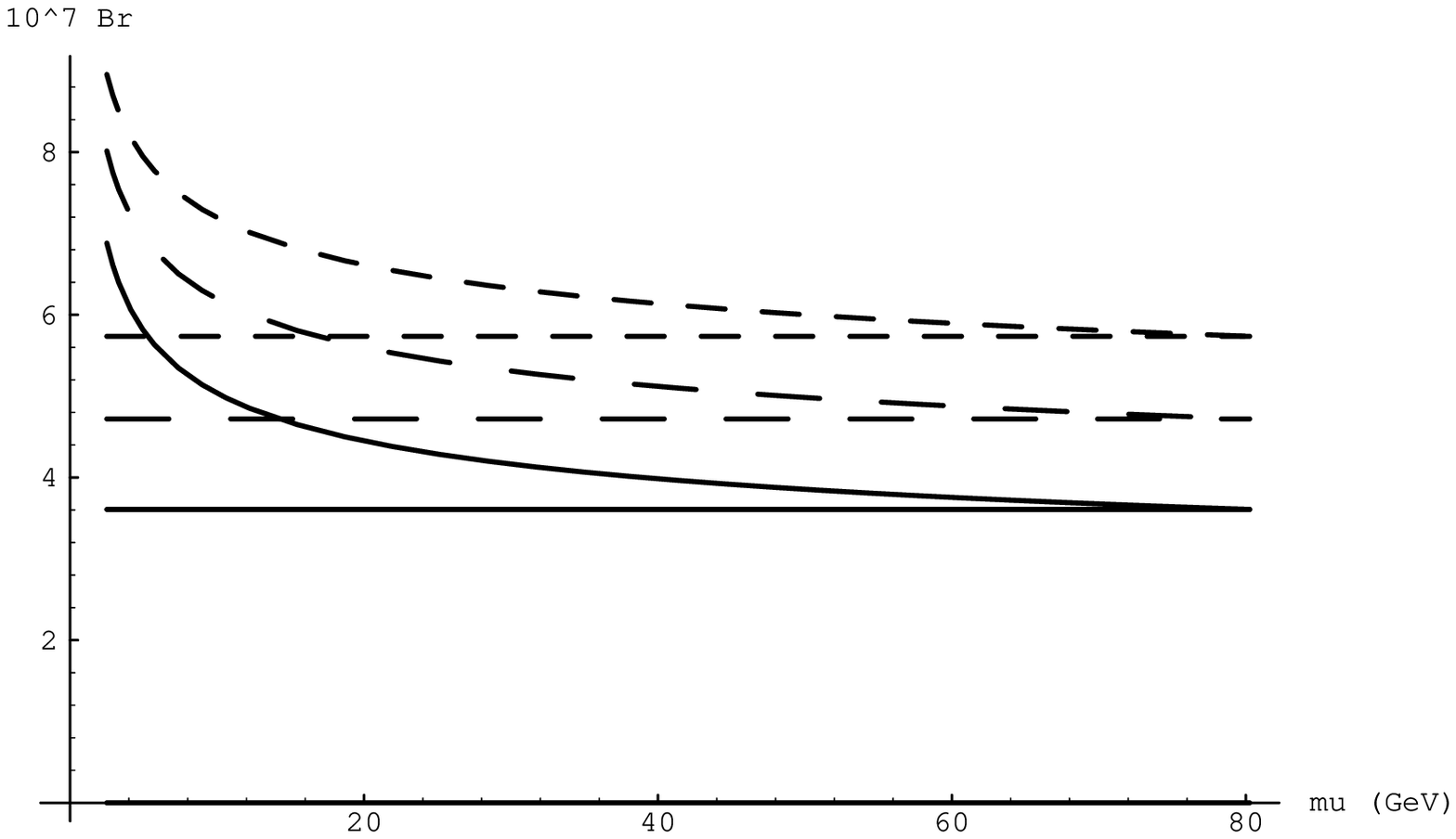}
\vskip -1.5truein
\caption[]{The scale dependence of the $Br$  
for set 3 and for 2 values of $m_{H}=500, \,\, 800\,\,GeV$ 
with $tan\beta=10$. 
Here, solid lines (curves) correspond to the SM
at $\mu=m_{W}$ (at arbitrary $\mu$ scale), dashed lines (curves) to 
model II 2HDM with $m_{H}=800\,\, GeV$ 
at $\mu=m_{W}$ (at arbitrary $\mu$ scale),
and small dashed lines (curves) model II 2HDM with $m_{H}=500\,\, GeV$ 
at $\mu=m_{W}$ (at arbitrary $\mu$ scale).}
\label{brmu10p3}
\end{figure}

\begin{figure}[htb]
\vskip -1.5truein
\centering
\epsfxsize=3.8in
\leavevmode\epsffile{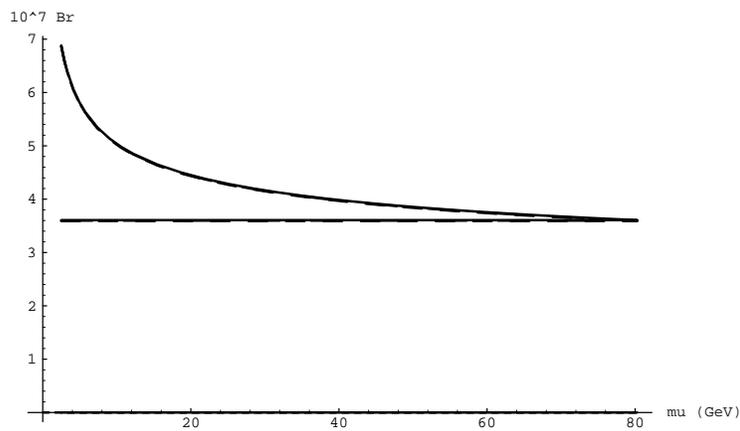}
\vskip -1.5truein
\caption[]{Same as fig~12, but for model I 2HDM. 
All curves coincide within errors.}
\label{brmu10Ip3}
\end{figure}

\begin{figure}[htb]
\vskip -1.5truein
\centering
\epsfxsize=3.8in
\leavevmode\epsffile{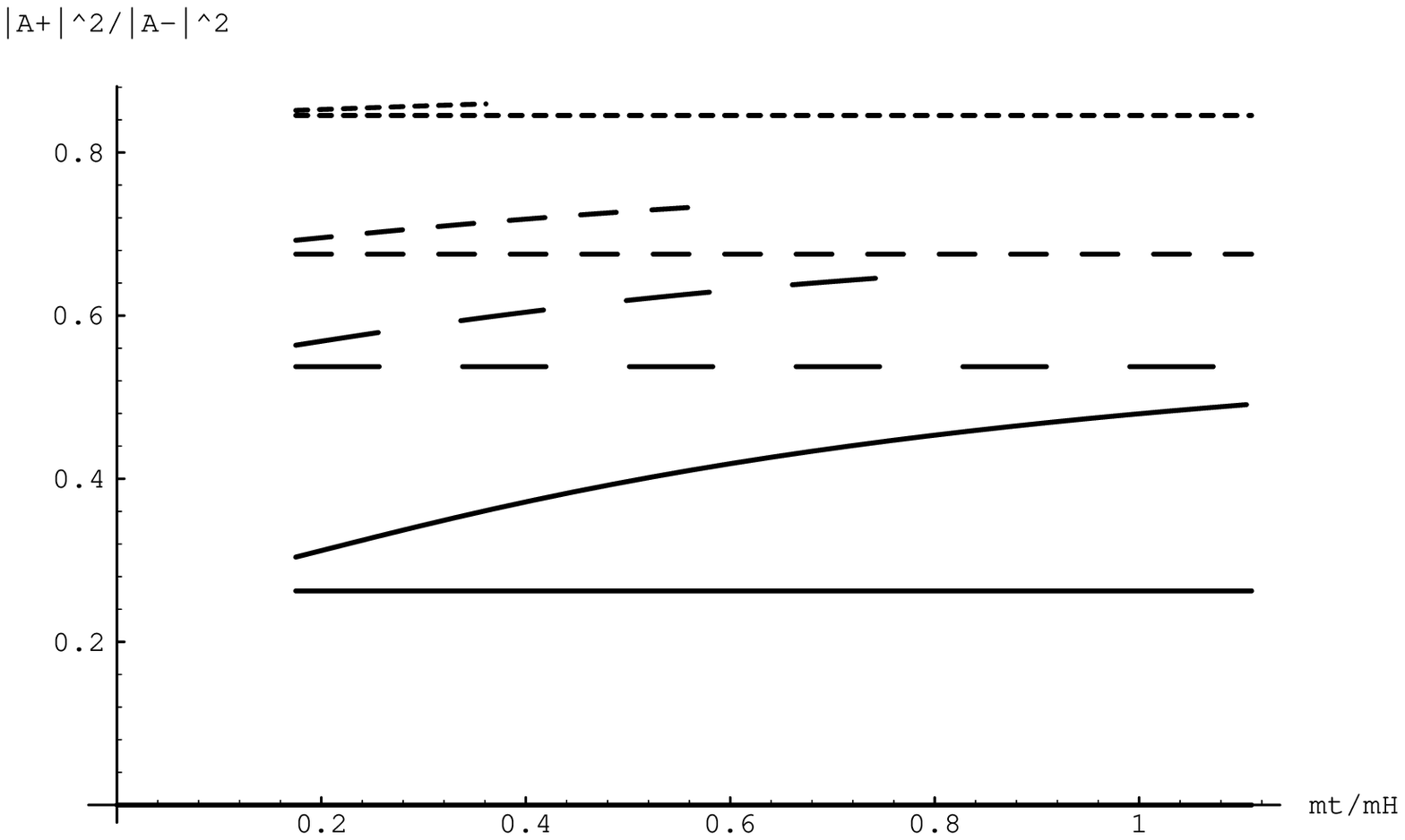}
\vskip -1.5truein
\caption[]{$m_{t}/m_{H}$ dependence of the ratio $|A^{+}|^2/|A^{-}|^2$
for set 3 at $tan\beta=10$ with the addition
of $LD_{O_{7}}$ effects. 
Here, solid lines (curves) correspond to the 
SM  (model II 2HDM) at $\mu=m_{W}$, long dashed lines (curves) to the 
SM (model II 2HDM) 
at $\mu=10\,\, GeV$, medium dashed lines (curves) to the SM (model II 2HDM) 
at $\mu=5\,\, GeV$ and small dashed lines (curves) to the SM (model II 2HDM) 
at $\mu=2.5\,\, GeV$.}
\label{ksi01p3LD}
\end{figure}

\begin{figure}[htb]
\centering
\vskip -1.5truein
\epsfxsize=3.8in
\leavevmode\epsffile{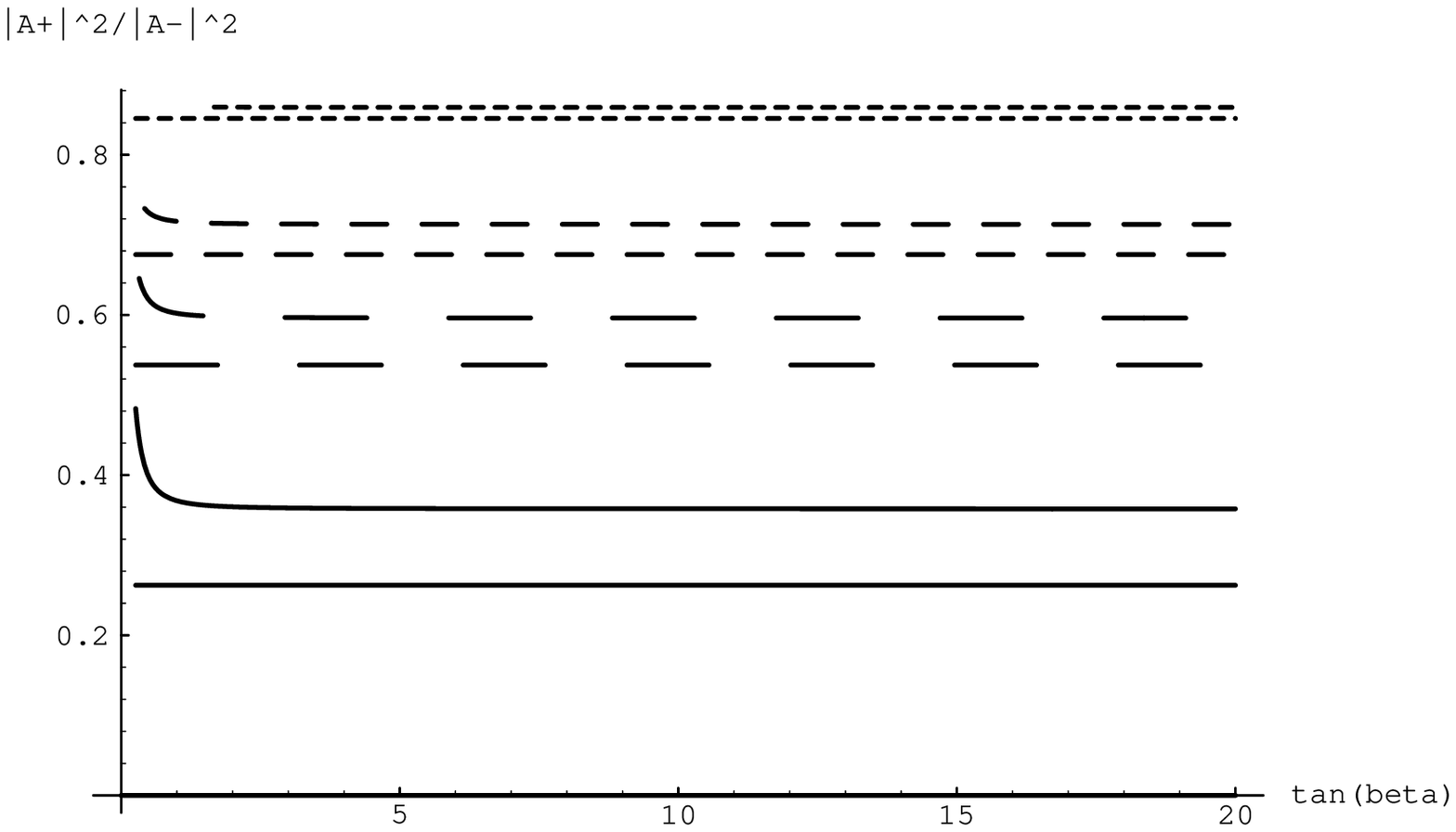}
\vskip -1.5truein
\caption[]{$tan\beta$ dependence of the ratio $|A^{+}|^2/|A^{-}|^2$
for set 3 for $m_{H}=500\,\, GeV$ 
with the addition of $LD_{O_{7}}$ effects. 
Here, solid lines (curves) correspond to the 
SM  (model II 2HDM) at $\mu=m_{W}$, long dashed lines (curves) to the 
SM (model II 2HDM) 
at $\mu=10\,\, GeV$, medium dashed lines (curves) to the SM (model II 2HDM) 
at $\mu=5\,\, GeV$ and small dashed lines (curves) to the SM (model II 2HDM) 
at $\mu=2.5\,\, GeV$.}
\label{yt500p3LD}
\end{figure}

\begin{figure}[htb]
\vskip -1.5truein
\centering
\epsfxsize=3.8in
\leavevmode\epsffile{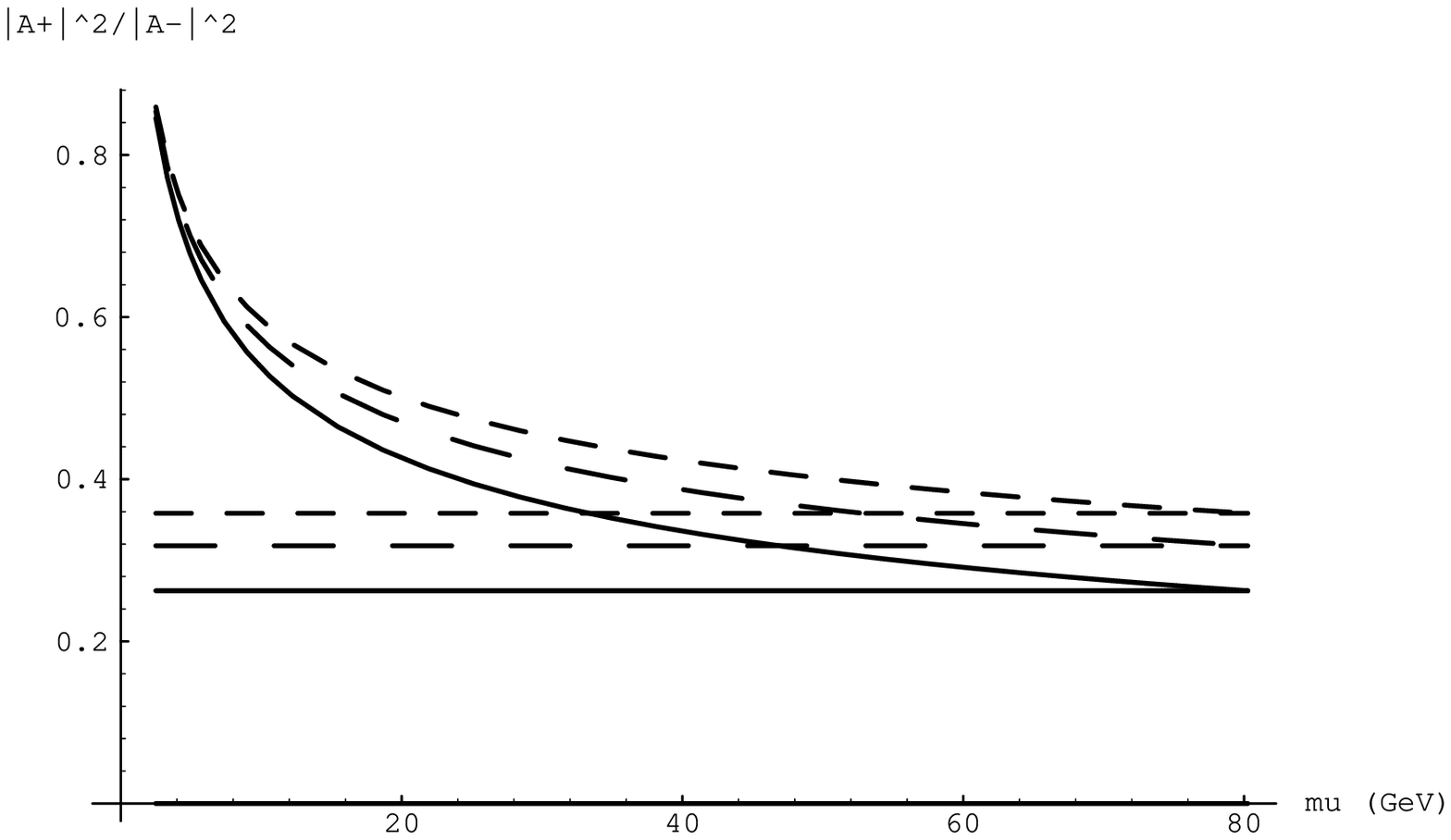}
\vskip -1.5truein
\caption[]{The scale dependence of the ratio $|A^{+}|^2/|A^{-}|^2$
for set 3 and for 2 values of $m_{H}=500,\,\, 800\,\,GeV$ 
at $tan\beta=10$ with the addition of $LD_{O_{7}}$ effects.
Here, solid lines (curves) correspond to the SM
at $\mu=m_{W}$ (at arbitrary $\mu$ scale), dashed lines (curves) to 
model II 2HDM with $m_{H}=800\,\, GeV$ 
at $\mu=m_{W}$ (at arbitrary $\mu$ scale),
and small dashed lines (curves) to model II 2HDM with $m_{H}=500\,\, GeV$ 
at $\mu=m_{W}$ (at arbitrary $\mu$ scale).}
\label{mu10p3LD}
\end{figure}

\begin{figure}[htb]
\vskip -1.5truein
\centering
\epsfxsize=3.8in
\leavevmode\epsffile{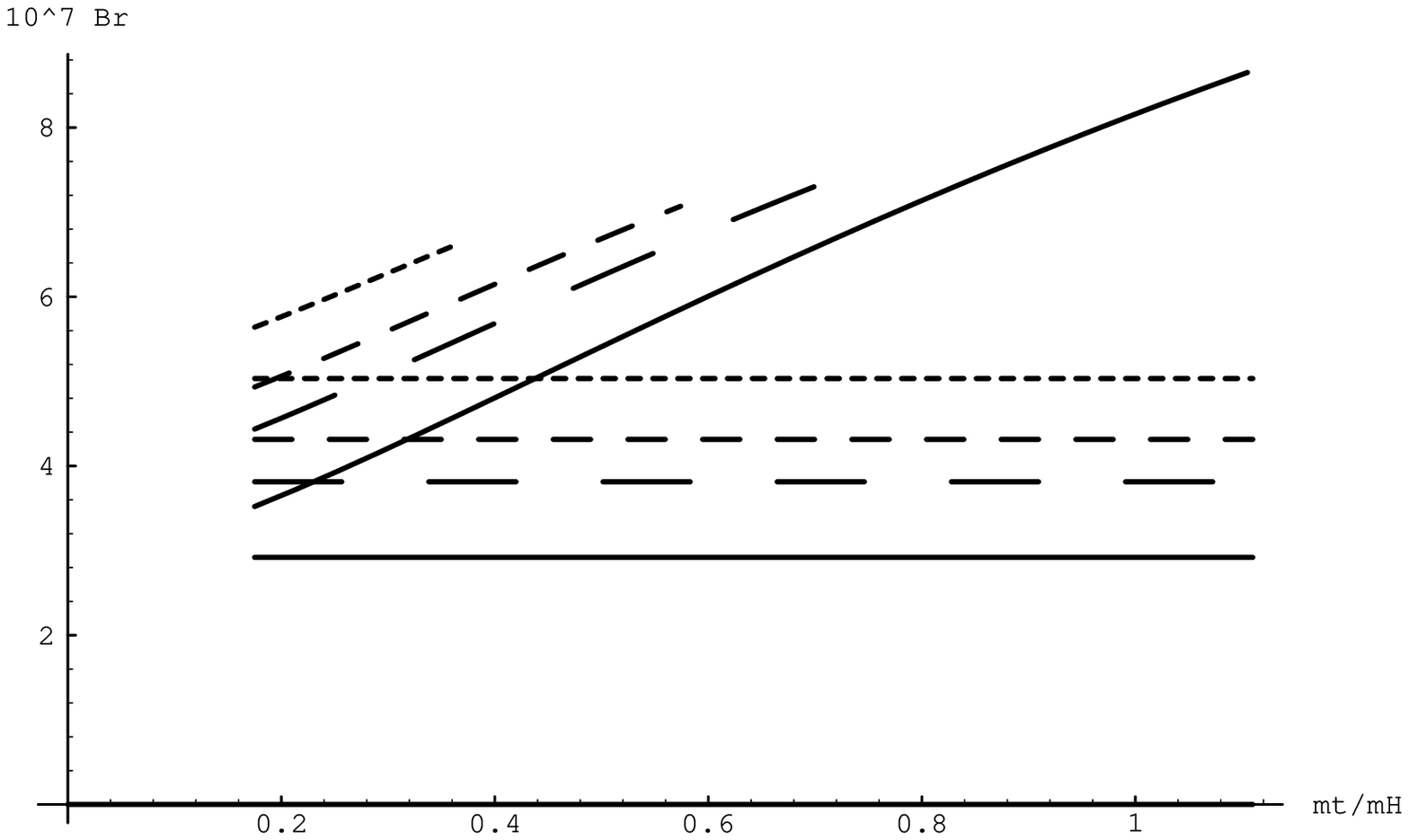}
\vskip -1.5truein
\caption[]{$m_{t}/m_{H}$ dependence of the $Br$  
for  set 3 at $tan\beta=10$ including $LD_{O_{7}}$ effects. 
Here, solid lines (curves) correspond to the
SM  (model II 2HDM) at $\mu=m_{W}$, long dashed lines (curves) to the 
SM (model II 2HDM) 
at $\mu=10\,\, GeV$, medium dashed lines (curves) to the SM (model II 2HDM) 
at $\mu=5\,\, GeV$ and small dashed lines (curves) to the SM (model II 2HDM) 
at $\mu=2.5\,\, GeV$.}
\label{brksi01p3LD}
\end{figure}

\begin{figure}[htb]
\centering
\vskip -1.5truein
\epsfxsize=3.8in
\leavevmode\epsffile{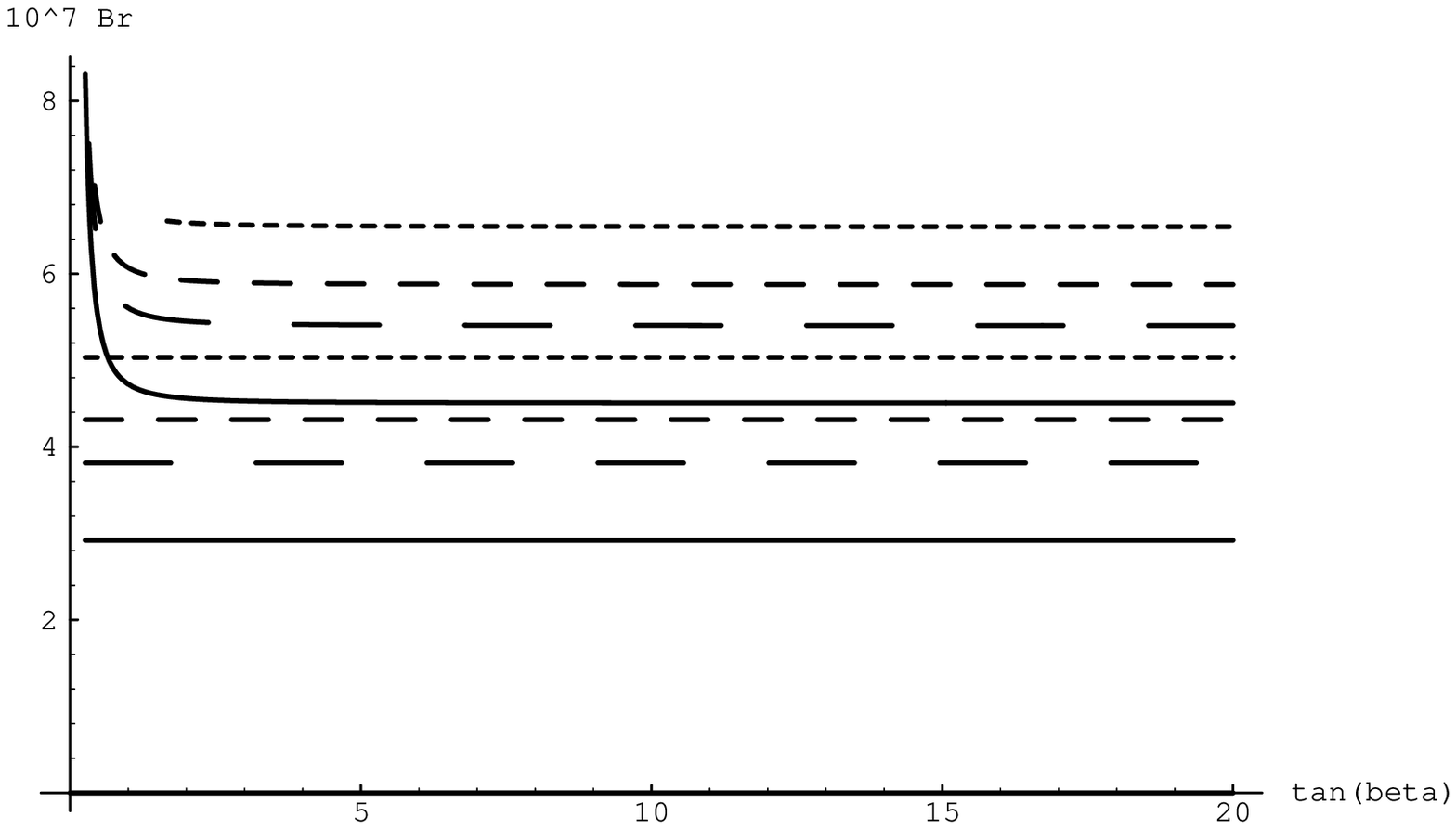}
\vskip -1.5truein
\caption[]{$tan\beta$ dependence of the $Br$  
for set 3 for $m_{H}=500\,\, GeV$  
including $LD_{O_{7}}$ effects. 
Here, solid lines (curves) correspond to the
SM  (model II 2HDM) at $\mu=m_{W}$, long dashed lines (curves) to the 
SM (model II 2HDM) 
at $\mu=10\,\, GeV$, medium dashed lines (curves) to the SM (model II 2HDM) 
at $\mu=5\,\, GeV$ and small dashed lines (curves) to the SM (model II 2HDM) 
at $\mu=2.5\,\, GeV$.}
\label{bryt500p3LD}
\end{figure}

\begin{figure}[htb]
\vskip -1.5truein
\centering
\epsfxsize=3.8in
\leavevmode\epsffile{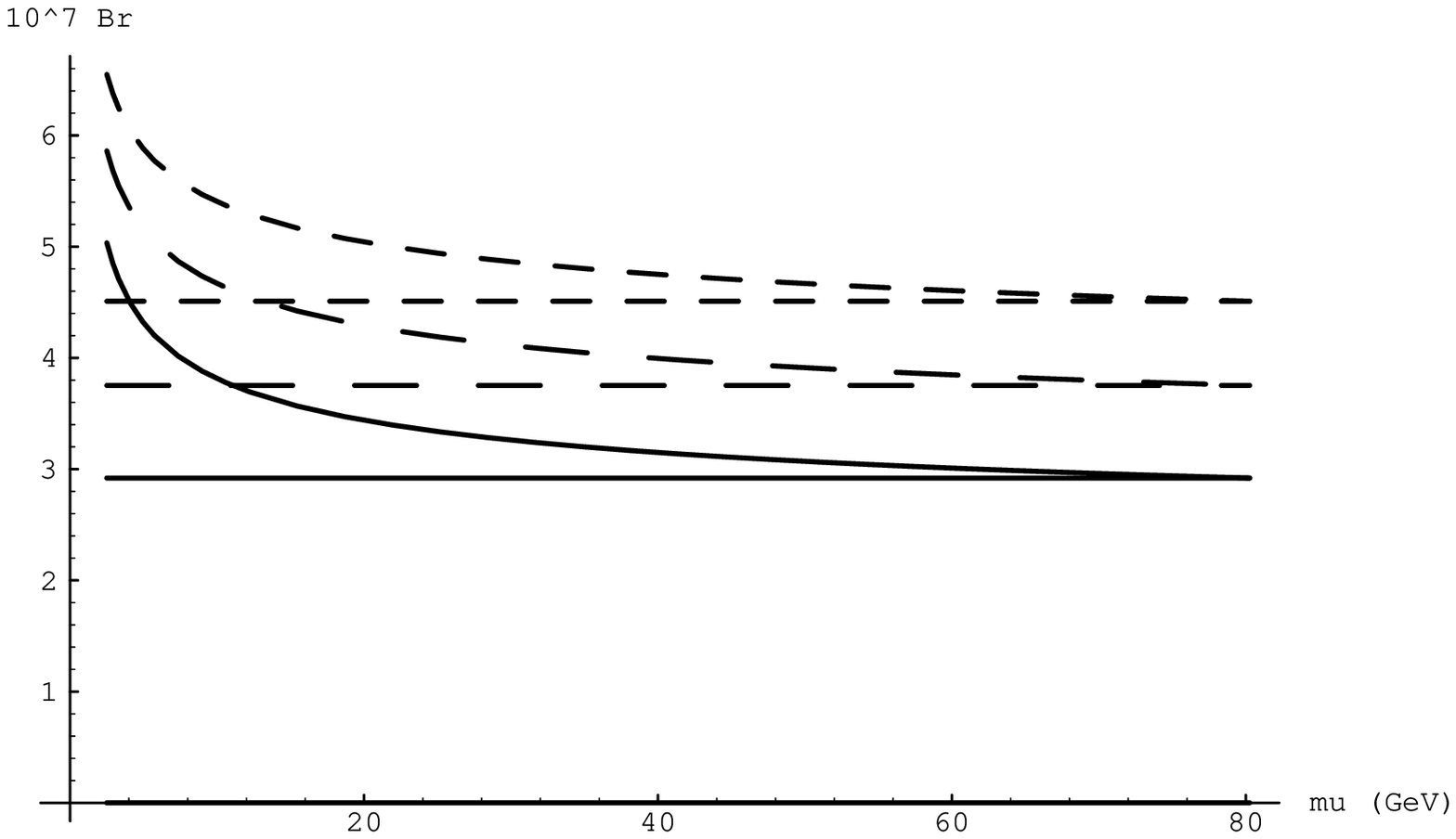}
\vskip -1.5truein
\caption[]{The scale dependence of the $Br$  
for set 3 and for 2 values of $m_{H}=500,\,\, 800\,\,GeV$ 
at $tan\beta=10$ including $LD_{O_{7}}$ effects.
Here, solid lines (curves) correspond to the SM
at $\mu=m_{W}$ (at arbitrary $\mu$ scale), dashed lines (curves) to 
model II 2HDM with $m_{H}=800\,\, GeV$ 
at $\mu=m_{W}$(at arbitrary $\mu$ scale)
and small dashed lines (curves) to model II 2HDM with $m_{H}=500\,\, GeV$ 
at $\mu=m_{W}$ (at arbitrary $\mu$ scale).}
\label{brmu10p3LD}
\end{figure}
\end{document}